\documentclass[11pt]{article}
\pdfoutput=1
\usepackage{jheppubmod}
\usepackage[punctsep]{collref}
\usepackage{graphicx}
\usepackage{amssymb}
\usepackage{amsmath,amssymb}
\usepackage{slashed}
\usepackage{hyperref}
\usepackage{caption}
\usepackage{xcolor}
\usepackage{dsfont}
\usepackage{verbatim}
\usepackage{subfig}
\usepackage{amsmath,amssymb,amscd,amsfonts,mathtools}
\usepackage{xcolor}
\definecolor{markgreen}{RGB}{230,243,230}
\definecolor{darkolivegreen}{rgb}{0.33, 0.42, 0.18}
\definecolor{darkpastelgreen}{rgb}{0.01, 0.75, 0.24}

\DeclareMathOperator{\arcsinh}{arcsinh}

\usepackage[toc,page]{appendix}
\usepackage{epsfig}
\usepackage{epstopdf}
\usepackage{latexsym}
\usepackage{graphicx}
\usepackage{booktabs}
\usepackage{bbm}
\usepackage{color}
\usepackage{physics}
\usepackage{tensor}
\usepackage{verbatim}
\usepackage{subfig}
\usepackage{tikz}
\usepackage{ifthen}
\usetikzlibrary{matrix}
\usetikzlibrary{decorations.markings,calc,shapes,decorations.pathmorphing,patterns,decorations.pathreplacing,arrows.meta}
\usetikzlibrary{positioning}
\usepackage{hyperref}
\usetikzlibrary{patterns}
\usetikzlibrary{positioning}

\pdfoutput=1
\makeatletter
\def\@fpheader{\relax}
\makeatother

\newboolean{showcomments}
\setboolean{showcomments}{false}
\newcommand\rem[1]{\ifthenelse{\boolean{showcomments}}{{#1}}{}}
\newcommand{\be}{\begin{equation}}
\newcommand{\ee}{\end{equation}}

\title{\Large Replica Wormholes and Entanglement Islands in the Karch-Randall Braneworld}
\author{Hao Geng}
\affiliation{Center for the Fundamental Laws of Nature, Harvard University, 17 Oxford St., Cambridge, MA, 02139, USA.}
\emailAdd{haogeng@fas.harvard.edu}
\abstract{The Karch-Randall braneworld provides a natural set-up to study the Hawking radiation from a black hole using holographic tools. Such a black hole lives on a brane and is highly quantum yet has a holographic dual as a higher dimensional classical theory that lives in the ambient space. Moreover, such a black hole is coupled to a nongravitational bath which is absorbing its Hawking radiation. This allows us to compute the entropy of the Hawking radiation by studying the bath using the quantum extremal surface prescription. The quantum extremal surface geometrizes into a Ryu-Takayanagi surface in the ambient space. The topological phase transition of the Ryu-Takayanagi surface in time from connecting different portions of the bath to the one connecting the bath and the brane gives the Page curve of the Hawking radiation that is consistent with unitarity. Nevertheless, there doesn't exit a derivation of the quantum extremal surface prescription and its geometrization in the Karch-Randall braneworld. In this paper, we fill this gap. We mainly focus on the case that the ambient space is (2+1)-dimensional for which explicit computations can be done in each description of the set-up. We show that the topological phase transition of the Ryu-Takayanagi surface corresponds to the formation of the replica wormhole on the Karch-Randall brane as the dominate contribution to the replica path integral. For higher dimensional situations, we show that the geometry of the brane satisfies Einstein's equation coupled with conformal matter. We comment on possible implications to the general rule of gravitational path integral from this equation.
}
\begin{document}
\maketitle
\flushbottom
\pagebreak
\section{Introduction}\label{sec:intro}
The Karch-Randall braneworld \cite{Karch:2000ct,Karch:2000gx} concerns the physics of a large family of non-fine-tuned branes embedded in an ambient AdS$_{d+1}$ spacetime. It nevertheless plays a key role in the recent progress of quantum gravity due to its doubly holographic nature \cite{Almheiri:2019hni,Almheiri:2019psy,Geng:2020qvw}. This allows explicit constructions of entanglement islands and the computation of the unitary Page curve of the Hawking radiation from black holes in any dimensions. In such computations both the black hole and the entanglement island live on the brane and the entanglement island overlaps the black hole interior. Furthermore, important lessons about quantum gravity has been learned from the Karch-Randall braneworld that entanglement islands can faithfully exist only in massive gravity theories. This is motivated by the observation that the graviton localized on the Karch-Randall braneworld is massive \cite{Geng:2020qvw} and has been proven to be generally true in any dimensions beyond the Karch-Randall braneworld \cite{Geng:2021hlu,Geng:2023zhq}.

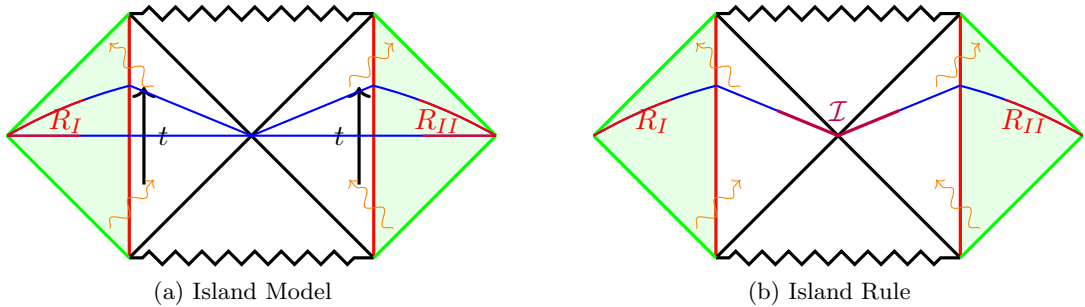
\begin{figure}[h]
    \centering
    \subfloat[Island Model \label{pic:penroseoriginal}]
{
    \begin{tikzpicture}[scale=0.65,decoration=snake]
       \draw[-,very thick] 
       decorate[decoration={zigzag,pre=lineto,pre length=5pt,post=lineto,post length=5pt}] {(-2.5,0) to (2.5,0)};
       \draw[-,very thick,red] (-2.5,0) to (-2.5,-5);
       \draw[-,very thick,red] (2.5,0) to (2.5,-5);
         \draw[-,very thick] 
       decorate[decoration={zigzag,pre=lineto,pre length=5pt,post=lineto,post length=5pt}] {(-2.5,-5) to (2.5,-5)};
       \draw[-,very thick] (-2.5,0) to (2.5,-5);
       \draw[-,very thick] (2.5,0) to (-2.5,-5);
       \draw[-,very thick,green] (-2.5,0) to (-5,-2.5);
       \draw[-,very thick,green] (-5,-2.5) to (-2.5,-5);
        \draw[-,very thick,green] (2.5,0) to (5,-2.5);
       \draw[-,very thick,green] (5,-2.5) to (2.5,-5);
       \draw[fill=green, draw=none, fill opacity = 0.1] (-2.5,0) to (-5,-2.5) to (-2.5,-5) to (-2.5,0);
       \draw[fill=green, draw=none, fill opacity = 0.1] (2.5,0) to (5,-2.5) to (2.5,-5) to (2.5,0);
       \draw[->,very thick,black] (-2.2,-3.5) to (-2.2,-1.5);
       \node at (-1.8,-2.5)
       {\textcolor{black}{$t$}};
        \draw[->,very thick,black] (2.2,-3.5) to (2.2,-1.5);
       \node at (1.8,-2.5)
       {\textcolor{black}{$t$}};
       \draw[-,thick, blue] (-5,-2.5) to (5,-2.5);
       \draw[-,thick,red] (-5,-2.5) to (-3.5,-2.5);
       \draw[-,thick,red] (5,-2.5) to (3.5,-2.5);
       \draw[-,thick,blue] (5,-2.5) arc (60:75.5:10);
       \draw[-,thick,blue] (-5,-2.5) arc (120:104.5:10);
       \draw[-,thick,blue] (2.5,-1.47) to (0,-2.5);
       \draw[-,thick,blue] (-2.5,-1.47) to (0,-2.5);
       \draw[-,thick,red] (5,-2.5) arc (60:70:10);
       \draw[-,thick,red] (-5,-2.5) arc (120:110:10);
       \node at (-3.8,-2.2)
       {\textcolor{red}{$R_{I}$}};
        \node at (3.8,-2.2)
       {\textcolor{red}{$R_{II}$}};
       \draw[->,decorate,orange] (-2.9,-4.4) to (-2,-3.4);
        \draw[->,decorate,orange] (-2,-1.5) to (-2.9,-0.6);
         \draw[->,decorate,orange] (2.9,-4.4) to (2,-3.4);
        \draw[->,decorate,orange] (2,-1.5) to (2.9,-0.6);
    \end{tikzpicture}}
    \hspace{1.0cm}
    \subfloat[Island Rule \label{pic:penroseisland}]
{
    \begin{tikzpicture}[scale=0.65,decoration=snake]
       \draw[-,very thick] 
       decorate[decoration={zigzag,pre=lineto,pre length=5pt,post=lineto,post length=5pt}] {(-2.5,0) to (2.5,0)};
       \draw[-,very thick,red] (-2.5,0) to (-2.5,-5);
       \draw[-,very thick,red] (2.5,0) to (2.5,-5);
         \draw[-,very thick] 
       decorate[decoration={zigzag,pre=lineto,pre length=5pt,post=lineto,post length=5pt}] {(-2.5,-5) to (2.5,-5)};
       \draw[-,very thick] (-2.5,0) to (2.5,-5);
       \draw[-,very thick] (2.5,0) to (-2.5,-5);
       \draw[-,very thick,green] (-2.5,0) to (-5,-2.5);
       \draw[-,very thick,green] (-5,-2.5) to (-2.5,-5);
        \draw[-,very thick,green] (2.5,0) to (5,-2.5);
       \draw[-,very thick,green] (5,-2.5) to (2.5,-5);
       \draw[fill=green, draw=none, fill opacity = 0.1] (-2.5,0) to (-5,-2.5) to (-2.5,-5) to (-2.5,0);
       \draw[fill=green, draw=none, fill opacity = 0.1] (2.5,0) to (5,-2.5) to (2.5,-5) to (2.5,0);
       \draw[-,thick,blue] (5,-2.5) arc (60:75.5:10);
       \draw[-,thick,blue] (-5,-2.5) arc (120:104.5:10);
       \draw[-,thick,blue] (2.5,-1.47) to (1.25,-1.985);
       \draw[-,very thick,purple!!!] (1.25,-1.985) to (0,-2.5);
       \draw[-,thick,blue] (-2.5,-1.47) to (-1.25,-1.985);
       \draw[-,very thick,purple!!!] (-1.25,-1.985) to (0,-2.5);
         \node at (0,-2.)
       {\textcolor{purple}{$\mathcal{I}$}};
       \draw[-,thick,red] (5,-2.5) arc (60:70:10);
       \draw[-,thick,red] (-5,-2.5) arc (120:110:10);
       \node at (-3.8,-2.2)
       {\textcolor{red}{$R_{I}$}};
        \node at (3.8,-2.2)
       {\textcolor{red}{$R_{II}$}};
       \draw[->,decorate,orange] (-2.9,-4.4) to (-2,-3.4);
        \draw[->,decorate,orange] (-2,-1.5) to (-2.9,-0.6);
         \draw[->,decorate,orange] (2.9,-4.4) to (2,-3.4);
        \draw[->,decorate,orange] (2,-1.5) to (2.9,-0.6);
    \end{tikzpicture}}
   
    \caption{\small 
    \textbf{a)} The Penrose diagram of an eternal black hole in AdS$_{d}$ coupled to $d$-dimensional baths. We specify the two red vertical lines as the conformal boundary of the AdS$_{d}$ black hole. The orange arrows are the radiation coming in and out of the black hole. We choose time evolution as indicated in the diagram. We also specify two Cauchy slices of this time evolution as the two blue curves and on each of them we denote the subsystem $R=R_{I}\cup R_{II}$ in red. We emphasize that the Cauchy slices of this time evolution all go through the bifurcation horizon so they don't touch the black interior. \textbf{b)} We draw a putative configuration with entanglement island as the purple interval in the black hole spacetime. Its causal diamond overlaps the black hole interior.}
\end{figure}

The aforementioned progress is based on the realization that Karch-Randall braneworld provides the holographic dual of the set-up in which the entropy of the black hole radiation can be consistently defined and hence calculated using holographic tools. This set-up contains a black hole in a gravitational asymptotically AdS$_{d}$ spacetime and is coupled to a nongravitational thermal bath on its asymptotic boundary (see Fig.\ref{pic:penroseoriginal}). The bath and the black hole are of the same temperature and in the thermal equilibrium. However, if we choose a proper time evolution (as indicated in Fig.\ref{pic:penroseoriginal}) then the dynamics of the black hole and the bath system is nontrivial in which the bath is collecting the radiation from the black hole. Hence we can compute the time evolution of the entanglement entropy of the early time Hawking radiation from the black hole by studying a subregion $R$ of the bath. The bath is nongravitational so such an entanglement entropy is well-defined \cite{Laddha:2020kvp,Raju:2020smc,Raju:2021lwh,Geng:2023qwm}. In such a computation, one is instructed to use the so called \textit{island rule} \cite{Almheiri:2019psf,Penington:2019npb} aka the \textit{quantum extremal surface prescription} \cite{Engelhardt:2014gca} to look for a quantum extremal surface $\partial I$
\begin{equation}
    S_{\text{vN}}(R)=\min_{\partial I} \text{Ext} \Big(S_{\text{vN}}(R\cup I)+\frac{A(\partial I)}{4G_{N}}\Big)\,,\label{eq:islandrule}
\end{equation}
in which $S_{\text{vN}}(R\cup I)$ is the entanglement (von-Neumann) entropy of the subregion $R\cup I$ for the quantum fields living in the black hole plus bath background, $G_{N}$ is the Newton's constant in the black hole spacetime and $\mathcal{I}$ is called the entanglement island. Using the above formula, one shall find the formation of a nontrivial island $\mathcal{I}$ at late times in our chosen time evolution. The interpretation of such a result is that at late times the entanglement wedge of the early-time Hawking radiation $R$ contains a disconnected region $\mathcal{I}$ from it. Entanglement wedge reconstruction tells us that the physics in the island region $\mathcal{I}$ is fully encoded in the early-time Hawking radiation $R$ \cite{Raju:2020smc,Almheiri:2020cfm}. Since the (causal diamond of the) island region overlaps with the black hole interior, one sees that the early-time Hawking radiation is indeed purified at late times as it encodes physics in the black hole interior. Nevertheless, the application of this island rule is extremely hard in generic situations as we don't have a formula for the subregion entanglement entropy for quantum fields in curved background in higher spacetime dimensions. This is the reason why Karch-Randall braneworld is a valuable set-up as the Karch-Randall braneworld provides a holographic model of the above set-up in Fig.\ref{pic:penroseoriginal}. In the Karch-Randall braneworld the quantum fields that live in the black hole plus bath background is holographic so we can in principle compute $S_{\text{vN}}(R\cup I)$ using holographic tools and explicitly construct entanglement islands using the island rule Equ.~(\ref{eq:islandrule}).

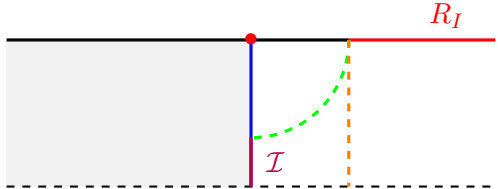
\begin{figure}[h]
    \centering
    \begin{tikzpicture}[scale=0.65]
         \draw[-,very thick](-5,0) to (2,0);
         \draw[-,very thick,red] (2,0) to (5,0);
         \draw[-,dashed,thick] (-5,-3) to (5,-3);
         \draw[-,very thick,blue] (0,0) to (0,-3);
         \node at (0,0)
         {\textcolor{red}{$\bullet$}};
         \node at (4,0.5)
         {\textcolor{red}{$R_{I}$}};
         \draw[fill=gray, draw=none, fill opacity = 0.1] (-5,0) to (0,0) to (0,-3) to (-5,-3);
         \draw[-,dashed,very thick,green] (2,0) arc(0:-90:2);
         \draw[-,dashed,very thick,orange] (2,0) to (2,-3);
           \draw[-,very thick,purple] (0,-2) to (0,-3);
         \node at (0.5,-2.5)
         {\textcolor{purple}{$\mathcal{I}$}};
    \end{tikzpicture}
    \caption{\small The demonstration of the Karch-Randall realization of the set-up in Fig.\ref{pic:penroseoriginal} on a constant time slice. For simplicity we only draw the configuration on the left hand side on the bulk eternal black hole. The dashed line is the horizon. The brane is the blue curve. The orange and the green curves are the two candidate entangling surfaces, that calculate $S_{\text{vN}}(R)=S_{\text{vN}}(R_{I}\cup R_{II})$, among which one goes through the bulk black hole interior connecting the boundaries of $R_{I}$ and $R_{II}$, and the other has two disconnected components one on each side of the bulk black hole and they end on the brane.}
    \label{pic:KRrealization}
\end{figure}

One can naturally realize the above set-up in Fig.\ref{pic:penroseoriginal} by embedding a Karch-Randall brane into an ambient eternal AdS$_{d+1}$ black hole background (see Fig.\ref{pic:KRrealization}). In this case, the ambient black hole will induce a black hole on the Karch-Randall brane which realizes the black hole in Fig.\ref{pic:penroseoriginal}. Moreover, the Karch-Randall brane cuts off part of the bulk (the gray region in Fig.\ref{pic:penroseoriginal}) and the leftover bulk region still has part of the asymptotic boundary which models the flat nongravitational bath in Fig.\ref{pic:penroseoriginal}. One usually takes the ambient space to be described by a classical Einstein's gravity (i.e. small curvature and small Newton's constant in the units that $l_{\text{AdS}}=1$). We call this the \textit{bulk description} of the Karch-Randall braneworld. Using the AdS/CFT correspondence \cite{Maldacena:1997re,Gubser:1998bc,Witten:1998qj}, one can firstly dualize the ambient space classical gravity theory to the brane plus the bath system that we now have a highly quantum (i.e. strongly coupled) gravitational theory with matter lives on the asymptotic AdS$_{d}$ black hole coupled to a nongravitational bath. This is exactly the set-up in Fig.\ref{pic:penroseoriginal}. Nevertheless, in our system, the bath is described by a d-dimensional conformal field theory (CFT$_{d}$) and the matter field on the AdS$_{d}$ black hole spacetime is a CFT$_{d}$ with a UV cutoff \cite{Geng:2023qwm}. We call this the \textit{intermediate description} of the Karch-Randall braneworld. Furthermore, one can dualize the resulting black hole plus bath system to a conformal field theory with boundary (BCFT$_{d}$) by dualizing the AdS$_{d}$ black hole to a CFT$_{d-1}$ living on the boundary of the bath CFT$_{d}$. In the case of an eternal black hole as we are considering, the resulting BCFT$_{d}$ system has two copies and is in the thermal-field-double-state (TFD). We call this the \textit{boundary description} of the Karch-Randall braneworld. In this description, the entanglement entropy of the early-time Hawking radiation is translated to the entanglement entropy of the BCFT$_{d}$ subregion $R$ when the BCFT$_{d}$ is in the time-evolved TFD. Then one can use the Ryu-Takayanagi conjecture in the AdS/BCFT correspondence \cite{Fujita:2011fp,Takayanagi:2011zk} to compute this entanglement entropy by looking for the Ryu-Takayanagi surface in the original ambient space (see Fig.\ref{pic:KRrealization}). There are two possible topologies of the Ryu-Takayanagi surface-- one has a single component going through the bulk black hole interior connecting the boundaries of the subregions $R_{I}$ and $R_{II}$ (the orange surface in Fig.\ref{pic:KRrealization}) and the other has two components connecting the boundaries of $R_{I}$ and $R_{II}$ respectively to points on the brane (see the purple surface in Fig.\ref{pic:KRrealization} for one of the components). The Ryu-Takayanagi surfaces are minimal area surfaces and for the surface of the second type one also has to minimize its area with respect to its ending point on the brane. The area of the surface of the first type is linearly growing in time and the area of the surface of the second type is constant in time. Therefore, the Hubeny-Rangamani-Takayanagi (HRT) formula \cite{Hubeny:2007xt} tells us that
\begin{equation}
    S_{\text{vN}}(R)=\min (\frac{A_{c}(t)}{4G_{d+1}},\frac{2A_{dc}}{4G_{d+1}})\,,\label{eq:RT}
\end{equation}
where $G_{d+1}$ is the Newton's constant in the ambient space and $A$ denotes the area of a connected component of a surface. It is obvious that at late times the surface of the second type will dominate which corresponds to the formation of entanglement island (see the purple region on the brane in Fig.\ref{pic:KRrealization}) and the minimization of the area over the ending points on the brane, when we are searching for the surface of the second type, corresponds to the same operation in the island rule Equ.~(\ref{eq:islandrule}). Hence, one can explicitly compute the Page curve of the black hole radiation and see the formation of entanglement island in the Karch-Randall braneworld \cite{Almheiri:2019hni,Almheiri:2019psy,Geng:2020qvw,Geng:2021mic,Chen:2020uac} (see \cite{Geng:2023qwm} for a review).

Nevertheless, both the island rule Equ.~(\ref{eq:islandrule}) in generic situations and the Ryu-Takayanagi conjecture in AdS/BCFT Equ.~(\ref{eq:RT}) lack a derivation. The island rule can be proven in specific low dimmensional models of the set-up in Fig.\ref{pic:penroseoriginal} which shows that the formation of entanglement island is due to the emergence of the replica wormhole as the dominating saddle in the path integral that computes $S_{vN}(R)$ \cite{Almheiri:2019qdq} (see \cite{Papadodimas:2013kwa,Penington:2019kki,Hartman:2020swn,Goto:2020wnk,Engelhardt:2020qpv,Belin:2020hea,Belin:2020jxr,Liu:2020gnp,Liu:2020jsv,Bousso:2020kmy,Pollack:2020gfa,Marolf:2020rpm,Marolf:2020xie,Anous:2021caj,Kawabata:2021vyo,Balasubramanian:2022gmo,Belin:2023efa,Climent:2024trz,Calmet:2024tgm} for other relevant works). Nevertheless, the observation that the formation of entanglement island is due to the emergence of replica wormhole is expected to be a general mechanism. The Ryu-Takayanagi conjecture can be proven in the ordinary AdS/CFT context by translating the boundary replica trick calculation of the entanglement entropy to a gravitational path integral on replicated spacetime manifolds \cite{Lewkowycz:2013nqa}. 

In this paper, we provide a derivation of the Ryu-Takayanagi conjecture in AdS/BCFT whose holographic dual is the quantum extremal surface prescription. We mainly focus on the case that the bulk description contains an eternal AdS$_{3}$ BTZ black hole \cite{Banados:1992wn} for which explicit computations of the entanglement entropy of $R$ can be performed in both the bulk and the boundary descriptions \cite{Sully:2020pza,Geng:2021iyq}. The bulk computation uses the Ryu-Takayanagi conjecture Equ.~(\ref{eq:RT}) and it gives exactly the same answer as the computation in the boundary description using BCFT$_{2}$ techniques. We prove the validity of the Ryu-Takayanagi conjecture in this case in the same spirit as \cite{Lewkowycz:2013nqa} by translating the replica computation of the entanglement entropy in the boundary description to a gravitational path integral in the bulk description using holographic dictionary \cite{Maldacena:1997re,Gubser:1998bc,Witten:1998qj}. Interestingly, we will see that the Ryu-Takayanagi conjecture Equ.~(\ref{eq:RT}) is indeed true in this case and the second class of surfaces correspond to the configuration that a single Karch-Randall brane connects the boundaries of different replica copies of the BCFT. This corresponds to the formation of replica wormhole in the intermediate description. We comments on generalizations of this derivation to cases with two Karch-Randall branes in the ambient BTZ black hole spacetime, which describes the information transfer between two black holes through a nongravitational bath \cite{Geng:2021iyq}, and higher dimensional situations. We will see that in the higher dimensional cases the brane obeys Einstein's equation coupled with conformal matter which circumvents the Witten-Yau theorem \cite{Witten:1999xp,Cai:1999dqz} for the replica wormhole to be an on-shell configuration in the replica gravitational path integral but provides nontrivial no-go theorem for the proposal from \cite{VanRaamsdonk:2020tlr}.

This paper is organized as following. In Sec.\ref{sec:review} we review relevant previous work both in the precise description of Karch-Randall branes and the computation of entanglement entropy when the ambient space is asymptotically AdS$_{3}$. In Sec.\ref{sec:rw} we provide a derivation of the Ryu-Takayanagi conjecture in AdS/BCFT when the bulk is asymptotically AdS$_{3}$ and show that the emergence of replica wormholes lies behind the formation of entanglement islands on the Karch-Randall brane. In Sec.\ref{sec:higherd} we comment on the generalization of the above derivation to higher dimensional cases. We conclude our paper with discussions in Sec.\ref{sec:conclusion}. We collect useful geometric results in the appendix.

\section{Review of Previous Work}\label{sec:review}
In this section, we review the precise description of the Karch-Randall braneworld and the computation of the entropy of Hawking radiation in the Karch-Randall braneworld using examples when the bulk is asymptotically AdS$_{3}$. Generalizations to higher dimensions are straightforward.

\subsection{The Physics of the Karch-Randall Braneworld}
Karch-Randall braneworld concerns the embedding of a subcritical brane in an ambient asymptotically AdS$_{d+1}$ spacetime $\mathcal{M}_{d+1}$. The physics in the ambient spacetime is described by classical Einstein's gravity and in which the brane is a codimension one hypersurface $\mathcal{M}_{d}$. The action of this system is given by
\begin{equation}
S=-\frac{1}{16\pi G_{d+1}}\int_{\mathcal{M}_{d+1}} d^{d+1}x\sqrt{-g} (R-2\Lambda)-\frac{1}{8\pi G_{d+1}}\int_{\mathcal{M}_{d}}d^{d}x\sqrt{-h}(K-T)\,,\label{eq:action1}
\end{equation}
where $K$ is the trace of the extrinsic curvature of the brane, $h_{\mu\nu}$ is the induced metric on the brane, $T$ is the tension of the brane and $\Lambda=-\frac{d(d-1)}{2l_{\text{AdS}}^2}$ (later we will set the AdS length scale $l_{\text{AdS}}=1$ for convenience) is the bulk cosmological constant. The boundary condition of the bulk metric is of the Neumann type near the boundary and the Dirichlet type near the leftover asymptotic boundary i.e. it has a nonzero fluctuation near the brane but zero fluctuation near the leftover asymptotic boundary. This corresponds to the fact that in the intermediate description of the Karch-Randall braneworld the brane is the gravitational AdS$_{d}$ spacetime and the leftover asymptotic boundary is the d-dimensional nongravitational bath. This system has two sets of equations of motion-- the Einstein's equation which corresponds to the vanishing of the bulk part of the variation of the action Equ.~(\ref{eq:action1}) and the brane embedding equation which corresponds to the brane part of the vanishing of the variation of the same action. The Einstein's equation determines the background geometry of the ambient spacetime and the brane embedding equation decides how the brane is embedded in the ambient spacetime and thus the induced geometry of the brane. The brane embedding equaiton is given by
\begin{equation}
    K_{\mu\nu}-Kh_{\mu\nu}=T h_{\mu\nu}\,.\label{eq:braneemb}
\end{equation}

Let's consider a few explicit examples when the ambient spacetime is (2+1)-dimensional to see how the above system operates and their physical interpretations relevant for our later discussions.

\subsubsection{Empty AdS$_{3}$ with One Brane}\label{sec:empty}
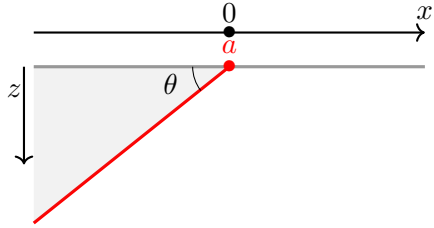
\begin{figure}[h]
\begin{centering}
\begin{tikzpicture}[scale=0.65]
\draw[-,very thick,black!40] (-4,0) to (0,0);
\draw[-,very thick,black!40] (0,0) to (4,0);
\draw[-,very thick,red] (0,0) to (-4,-3.2);
\draw[fill=gray, draw=none, fill opacity = 0.1] (0,0) to (-4,-3.2) to (-4,0) to (0,0);
\draw[-] (-0.75,0) arc (180:217.5:0.8);
\node at (0,0.4) {\textcolor{red}{$a$}};
\node at (-1.2,-0.35) {$\theta$};
\node at (0,0.7) {$\bullet$};
\node at (0,1.1) {$0$};
\node at (0,0) {\textcolor{red}{$\bullet$}};
\draw[->,thick,color=black] (-4,0.7) to (4,0.7);
\node at (4,1.1)
{\textcolor{black}{$x$}};
\draw[->,thick,color=black] (-4.2,0) to (-4.2,-2);
\node at (-4.4,-0.5)
{\textcolor{black}{$z$}};
\end{tikzpicture}
\caption{\small The brane configuration in an empty AdS$_{3}$. The configuration is static so we just draw a constant time slice. The brane subtend an angle $\theta$ with the asymptotic boundary $z=0$. The gray region behind the brane (for which $x<0$) is cut off by the brane.}
\label{pic:emptyads3}
\end{centering}
\end{figure}

As a first example, let's consider the simplest case that the bulk ambient space doesn't contain a black hole. In this case, the bulk geometry can be taken as in the Poincar\'{e} patch
\begin{equation}
    ds^{2}=\frac{dz^2-dt^2+dx^2}{z^2}\,,
\end{equation}
which solves the Einstein's equation for $d=2$ and $z\rightarrow0$ is the asymptotic boundary. The static solution of the brane embedding equation Equ.~(\ref{eq:braneemb}) is (see Fig.\ref{pic:emptyads3})
\begin{equation}
z=-x\tan\theta\,,\quad\text{where $T=\cos\theta$}\,.
\end{equation}

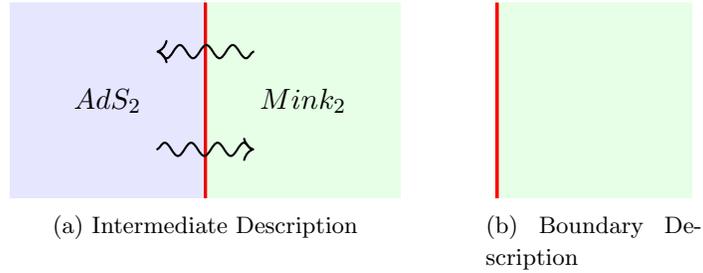
\begin{figure}[h]
    \centering
    \subfloat[Intermediate Description\label{pic:emptyads3intermediate}]{\begin{tikzpicture}[scale=0.65]
      \draw[-,very thick,red](0,-2) to (0,2);
       \draw[fill=green, draw=none, fill opacity = 0.1] (0,-2) to (4,-2) to (4,2) to (0,2);
           \draw[-,very thick,red](0,-2) to (0,2);
       \draw[fill=blue, draw=none, fill opacity = 0.1] (0,-2) to (-4,-2) to (-4,2) to (0,2);
       \node at (-2,0)
       {\textcolor{black}{$AdS_{2}$}};
        \node at (2,0)
       {\textcolor{black}{$Mink_{2}$}};
       \draw [-{Computer Modern Rightarrow[scale=1.25]},thick,decorate,decoration=snake] (-1,-1) -- (1,-1);
       \draw [-{Computer Modern Rightarrow[scale=1.25]},thick,decorate,decoration=snake] (1,1) -- (-1,1);
    \end{tikzpicture}}
    \hspace{1cm}
    \subfloat[Boundary Description\label{pic:emptyads3boundary}]{
\begin{tikzpicture}[scale=0.65]
      \draw[-,very thick,red](0,-2) to (0,2);
       \draw[fill=green, draw=none, fill opacity = 0.1] (0,-2) to (4,-2) to (4,2) to (0,2);
           \draw[-,very thick,red](0,-2) to (0,2);
    \end{tikzpicture}
    }
    \caption{\small \textbf{a)} Intermediate description: Gravitational AdS$_{2}$ coupled with a nongravitational (1+1)-dimensional half-Minkowski bath. \textbf{b)} Boundary description: a BCFT$_{2}$ in the ground state.}
\end{figure}

The intermediate description of this case is that we have an Poincar\'{e} AdS$_{2}$ gravitatonal spacetime coupled with a two-dimensional half-Minkowski nongravitaitonal bath. The matter field in the union of AdS$_{2}$ and the bath obeys transparent boundary condition near their common boundary and is in the ground state (see Fig.\ref{pic:emptyads3intermediate}). The boundary description is a (1+1)-dimensional boundary conformal field theory (BCFT$_{2}$) in the ground state. This BCFT$_{2}$ lives on a half-Minkowski spacetime and its boundary is described by a Cardy boundary state $\ket{B_{a}}$ \cite{Cardy:2017ufe} whose boundary entropy is given by \cite{Fujita:2011fp}
\begin{equation}
    \ln g_{a} =\cot\frac{\theta}{2}=\sqrt{\frac{1+T}{1-T}}.
\end{equation}

\subsubsection{BTZ Black Hole with One Brane}\label{sec:BTZ1}
\begin{figure}[h]
\begin{centering}
\begin{tikzpicture}[scale=0.65]
\draw[-,very thick,blue!40] (-6,0) to (4,0);
\node at (-2,0) {\textcolor{red}{$\bullet$}};
\node at (-2,0.4) {\textcolor{red}{$a$}};
\draw[-,dashed,very thick,black!40] (-6,-4) to (4,-4);
\draw[-,very thick,red] (-2,0) arc (100:175:4.4); 
\draw[fill=gray, draw=none, fill opacity = 0.1] (-6,0)--(-2,0) arc (100:175:4.4)--(-6,-4) ;
\draw[->,thick,black] (-6,0.7) to (4,0.7);
\node at (-2,0.7) {\textcolor{black}{$\bullet$}};
\node at (-2,1.1) {\textcolor{black}{$0$}};
\node at (4,1.1) {\textcolor{black}{$x$}};
\draw[->, thick,black] (-6, -0.1) to (-6, -3.5);
\node at (-6.4,-3.5) {\textcolor{black}{$z$}};
\end{tikzpicture}
\caption{\small The brane configuration in an eternal BTZ black hole background. The configuration is static so we just draw a constant time slice and for simplicity we only draw one exterior. The dashed line is the black hole horizon $z=z_{h}$. The gray region behind the brane is cut off by the brane.}
\label{pic:AdS3BTZ1}
\end{centering}
\end{figure}
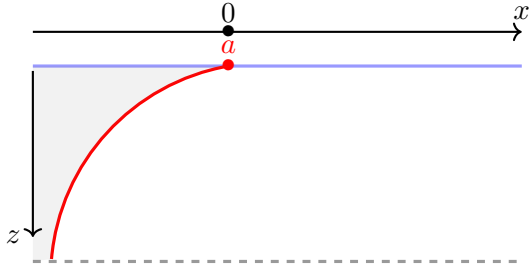

Now let's consider the case where the bulk ambient spacetime contains an eternal BTZ black hole. The metric is
    \begin{equation}
 ds^2=-\frac{h(z)}{z^2}dt^2+\frac{dz^2}{h(z)z^2}+\frac{dx^2}{z^2} \,,\qquad h(z)=1-\frac{z^2}{z_{h}^2} \,,  
\end{equation}
where the horizon is at $z=z_{h}$. The static solution of the brane embedding equation Equ.~(\ref{eq:braneemb}) is (see Fig.\ref{pic:AdS3BTZ1} and for details see \cite{Geng:2021iyq}\footnote{See \cite{Yu:2024fks,Feng:2024uia,Tan:2024url,Myers:2024zhb,Bernamonti:2024fgx,Liu:2024cmv,Deddo:2023oxn,Lin:2023hzs,Aguilar-Gutierrez:2023ccv,Yadav:2023qfg,Yadav:2023sdg,Basak:2023bnc,Liu:2023ggg,Jian:2023mdh,Kehrein:2023yeu,Basu:2023jtf,Ageev:2023hxe,Xu:2023fad,Deng:2023pjs,Kim:2023adq,Murdia:2023xvk,Chua:2023ios,Li:2023fzx,Balasubramanian:2023xyd,Mori:2023axa,Chang:2023gkt,Mori:2023swn,Chou:2023adi,Afrasiar:2023nir,Yu:2023whl,Jeong:2023lkc,Tan:2023cra,RoyChowdhury:2023eol,Karch:2023ekf,Tian:2023vbi,Afrasiar:2023jrj,Perez-Pardavila:2023rdz,Emparan:2023dxm,Baek:2022ozg,Afrasiar:2022fid,Bhattacharjee:2022pcb,Pasquarella:2022ibb,Lu:2022tmt,Craps:2022ahp,Antonini:2022sfm,Biswas:2022xfw,Geng:2022dua,HosseiniMansoori:2022hok,Ageev:2022qxv,Miyaji:2022dna,Yu:2022xlh,Geng:2022tfc,Bissi:2022bgu,Lee:2022efh,Kusuki:2022wns,Afrasiar:2022ebi,Basu:2022reu,Tian:2022pso,BasakKumar:2022stg,Bianchi:2022ulu,Kawamoto:2022etl,Lin:2022aqf,Numasawa:2022cni,Grimaldi:2022suv,Martinec:2022ofs,Wang:2021xih,Kusuki:2021gpt,Geng:2021mic,Ali:2021jcr,Bhattacharya:2021nqj,Omidi:2021opl,Collier:2021ngi,Chou:2021boq,Miyata:2021qsm,Uhlemann:2021itz,Iizuka:2021tut,Kibe:2021gtw,Shaghoulian:2021cef,He:2021mst,Belin:2021nck,Arefeva:2021kfx,Bhattacharya:2021dnd,Hollowood:2021wkw,Kastikainen:2021ybu,Sun:2021dfl,Hernandez:2021goj,Balasubramanian:2021xcm,Pedraza:2021cvx,Ageev:2021ipd,Ahn:2021chg,Belin:2021htw,Caceres:2021fuw,KumarBasak:2021rrx,Akal:2021foz,Lu:2021gmv,Chu:2021gdb,Neuenfeld:2021bsb,Uhlemann:2021nhu,Balasubramanian:2021wgd,Hollowood:2021nlo,Miyata:2021ncm} for relevant follow-up works on \cite{Geng:2021iyq}.})
\begin{equation}
    x(z)=- z_h \arcsinh{\left(\frac{z\,T}{z_{h}\sqrt{1-T^2}}\right)} \,.
\end{equation}

The intermediate description in this case is the same as Fig.\ref{pic:penroseoriginal} where now the black hole is (1+1)-dimensional. The boundary description involves the same BCFT$_{2}$ with the boundary condition corresponds to the Cardy state $\ket{B_{a}}$ as in Sec.\ref{sec:empty}. But now there are two copies of such BCFT$_{2}$'s and they are in the thermal-field-double-state $\ket{\text{TFD}}$ (more details will be discussed in Sec.\ref{sec:boundary}).

\subsubsection{BTZ Black Hole with Two Branes}\label{sec:BTZ2}
\begin{figure}[h]
\begin{centering}
\begin{tikzpicture}[scale=0.65]
\draw[-,very thick,blue!40] (-6,0) to (6,0);
\node at (-2,0) {\textcolor{red}{$\bullet$}};
\node at (-2,0.4) {\textcolor{red}{$a$}};
\node at (2,0) {\textcolor{orange}{$\bullet$}};
\node at (2,0.4) {\textcolor{orange}{$b$}};
\draw[-,dashed,very thick,black!40] (-6,-4) to (6,-4);
\draw[-,very thick,red] (-2,0) arc (100:175:4.4); 
\draw[-,very thick,orange] (2,0) arc (50:-9:4.4); 
\draw[fill=gray, draw=none, fill opacity = 0.1] (-6,0)--(-2,0) arc (100:175:4.4)--(-6,-4) ;
\draw[fill=gray, draw=none, fill opacity = 0.1] (6,0)--(2,0) arc (50:-9:4.4)--(6,-4) ;
\end{tikzpicture}
\caption{\small The configuration with two branes in the BTZ background. The shaded gray region is cut off by the branes. }
\label{pic:AdS3BTZ2}
\end{centering}
\end{figure}
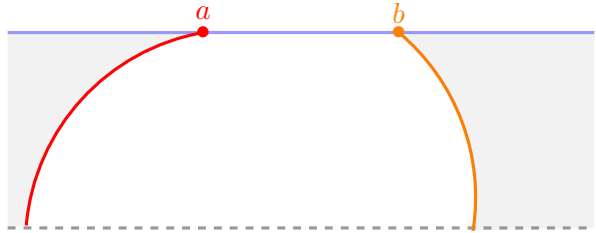

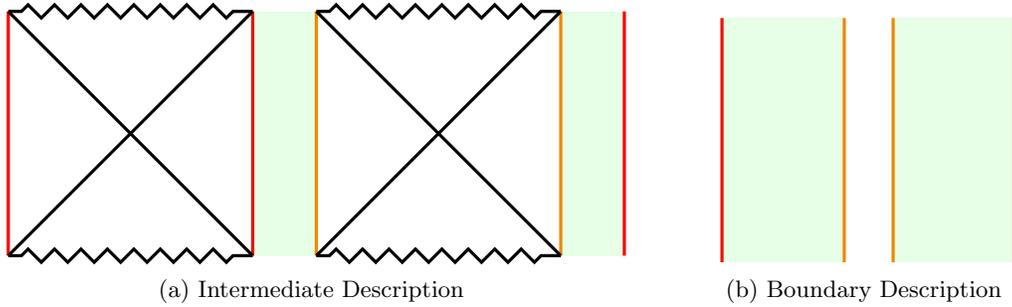
\begin{figure}[h]
    \centering
    \subfloat[Intermediate Description\label{pic:BTZ2intermediate}]{
    \begin{tikzpicture}[scale=0.65]
       \draw[-,very thick] 
       decorate[decoration={zigzag,pre=lineto,pre length=5pt,post=lineto,post length=5pt}] {(-2.5,0) to (2.5,0)};
       \draw[-,very thick,red] (-2.5,0) to (-2.5,-5);
       \draw[-,very thick,red] (2.5,0) to (2.5,-5);
         \draw[-,very thick] 
       decorate[decoration={zigzag,pre=lineto,pre length=5pt,post=lineto,post length=5pt}] {(-2.5,-5) to (2.5,-5)};
       \draw[-,very thick] (-2.5,0) to (2.5,-5);
       \draw[-,very thick] (2.5,0) to (-2.5,-5);
          \draw[-,very thick] 
       decorate[decoration={zigzag,pre=lineto,pre length=5pt,post=lineto,post length=5pt}] {(3.8,0) to (8.8,0)};
       \draw[-,very thick,orange] (3.8,0) to (3.8,-5);
       \draw[-,very thick,orange] (8.8,0) to (8.8,-5);
         \draw[-,very thick] 
       decorate[decoration={zigzag,pre=lineto,pre length=5pt,post=lineto,post length=5pt}] {(3.8,-5) to (8.8,-5)};
       \draw[-,very thick] (3.8,0) to (8.8,-5);
       \draw[-,very thick] (3.8,-5) to (8.8,0);
                \draw[fill=green, draw=none, fill opacity = 0.1] (2.5,0) to (2.5,-5) to (3.8,-5) to (3.8,0);
                \draw[fill=green, draw=none, fill opacity = 0.1] (8.8,0) to (8.8,-5) to (10.1,-5) to (10.1,0);
        \draw[-,very thick,red] (10.1,-5) to (10.1,0);
    \end{tikzpicture}}
\hspace{1cm}
    \subfloat[Boundary Description\label{pic:BTZ2bdy}]{
 \begin{tikzpicture}[scale=0.65]
       \draw[-,very thick,red] (2.5,0) to (2.5,-5);
       \draw[-,very thick,orange] (5,0) to (5,-5);
       \draw[-,very thick,orange] (6,0) to (6,-5);
                \draw[fill=green, draw=none, fill opacity = 0.1] (2.5,0) to (2.5,-5) to (5,-5) to (5,0);
                \draw[fill=green, draw=none, fill opacity = 0.1] (6,0) to (6,-5) to (8.5,-5) to (8.5,0);
        \draw[-,very thick,red] (8.5,-5) to (8.5,0);
    \end{tikzpicture}}
    \caption{\small \textbf{a)} The Penrose diagram of the intermediate description for two branes in the BTZ black hole background. We have two (1+1)-dimensional black holes (i.e the geometry on the two branes) coupled to each other through strip-shaped baths (the green shaded region). The two exterior red edges should be identified. We notice that the black hole singularities are in fact orbifold singularities inheriting from that of the BTZ black hole but we still use waved lines to represent them. \textbf{b)} The boundary description of the configuration with two branes in the BTZ black hole background. The two strip BCFT$_{2}$'s are in the thermal-field-double-state.}   
\end{figure}

A slight generalization of the situation in Sec.\ref{sec:BTZ1} is to put two Karch-Randall branes in the ambient BTZ black hole spacetime (see Fig.\ref{pic:AdS3BTZ2}). In this case, the intermediate description contains two (1+1)-dimensional eternal black holes coupled to each other through strip-shaped baths (see Fig.\ref{pic:BTZ2intermediate}). The boundary description involves two BCFT$_{2}$'s living on strips with boundary conditions corresponding to Cardy states $\ket{B_{a}}$ and $\ket{B_{b}}$. The two BCFT$_{2}$'s are in the thermal-field-double-state $\ket{\text{TFD}}$ (see Fig.\ref{pic:BTZ2bdy} and more details will be discussed in Sec.\ref{sec:boundary}).

\subsection{Computing the Entropy of Hawking Radiation}
With the understanding that the entropy of Hawking radiation is in fact the entanglement entropy of the bath subregion $R$ in Fig.\ref{pic:penroseoriginal}, we can try computing it in the Karch-Randall braneworld. The nice feature of the case that the ambient bulk spacetime is (2+1)-dimensional is that the computation can be done in both the boundary description using BCFT$_2$ techniques and the bulk description using the Ryu-Takayanagi conjecture Equ.~(\ref{eq:RT}) with exactly matched answers. We will review this computation in various cases in this subsection to set up some notations for our later discussion.

\subsubsection{General Idea}\label{sec:replicatrick}
In quantum field theories, for states that can be prepared using path integral, subregion entanglement entropy can be computed using the replica trick \cite{Solodukhin:2011gn}. Let's review this idea using the ground state $\ket{0}$ of a BCFT$_{2}$ on the plane with a single Cardy boundary. The state $\ket{0}$ is supported on the zero time slice and we are interested in the ground state entanglement entropy between the subregion $R$, which contains the boundary, and its complement $\bar{R}$. 

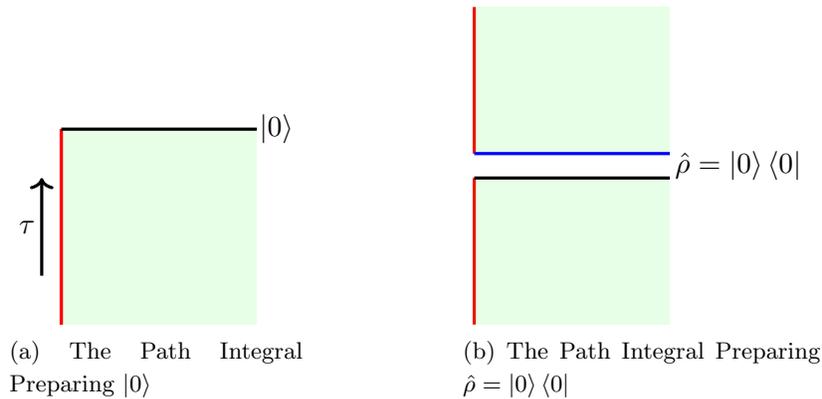
\begin{figure}[h]
    \centering
    \subfloat[The Path Integral Preparing $\ket{0}$ \label{pic:prepare0}]{\begin{tikzpicture}[scale=0.65]
      \draw[-,very thick,red](0,-2) to (0,2);
       \draw[fill=green, draw=none, fill opacity = 0.1] (0,-2) to (4,-2) to (4,2) to (0,2);
           \draw[-,very thick,red](0,-2) to (0,2);
           \draw[->,very thick,black] (-0.4,-1) to (-0.4,1);
           \node at (-0.7,0) {\textcolor{black}{$\tau$}};
           \draw[-,very thick,black] (0,2) to (4,2);
           \node at (4.4,2) {\textcolor{black}{$\ket{0}$}};
    \end{tikzpicture}}
    \hspace{2cm}
    \subfloat[The Path Integral Preparing $\hat{\rho}=\ket{0}\bra{0}$ \label{pic:preparerho}]{
\begin{tikzpicture}[scale=0.65]
      \draw[-,very thick,red](0,-3) to (0,0);
      \draw[-,very thick,red](0,0.5) to (0,3.5);
       \draw[fill=green, draw=none, fill opacity = 0.1] (0,-3) to (4,-3) to (4,0) to (0,0);
       \draw[fill=green, draw=none, fill opacity = 0.1] (0,0.5) to (4,0.5) to (4,3.5) to (0,3.5);
       \draw[-,very thick,black] (0,0) to (4,0);
       \draw[-,very thick,blue] (0,0.5) to (4,0.5);
       \node at (5.4,0.25) {\textcolor{black}{$\hat{\rho}=\ket{0}\bra{0}$}};
    \end{tikzpicture}
    }
    \caption{\small \textbf{a)} The Euclidean path integral preparing the ground state $\ket{0}$. The black slice is the $\tau=0$ slice where the state is supported. We can specify boundary values of fields $\phi(x)$ along the black slice if we want to compute the ground state wavefunctional $\Psi_{gs}(\phi)=\bra{\phi}\ket{0}$. \textbf{b)} The Euclidean path integral preparing the ground state density matrix. If one is interested in the specific element $\bra{\phi_{1}}\hat{\rho}\ket{\phi_{2}}$ then one can specify the values of the fields $\phi^{*}_{2}(x)$ on the blue slice and $\phi_{1}(x)$ on the black slice.}
\end{figure}

The ground state can be prepared using Euclidean time evolution from $-\infty$ to zero which projects all states with a nonzero overlap with the ground state to the ground state (see Fig.\ref{pic:prepare0}). Hence we will consider the BCFT$_{2}$ to be living on a Euclidean half-plane
\begin{equation}
    ds^2=d\tau^2+dx^2=dzd\bar{z}\,,\quad\text{where } z=\tau+ix\quad \text{and $x\geq 0$},\label{eq:plane}
\end{equation}
for which $x=0$ is the conformal boundary. Instead of considering the ground state $\ket{0}$, let's consider the corresponding density matrix operator
\begin{equation}
    \hat{\rho}=\ket{0}\bra{0}\,,\label{eq:rho0}
\end{equation}
which can again be prepared by a Euclidean path integral as indicated in Fig.\ref{pic:preparerho}.

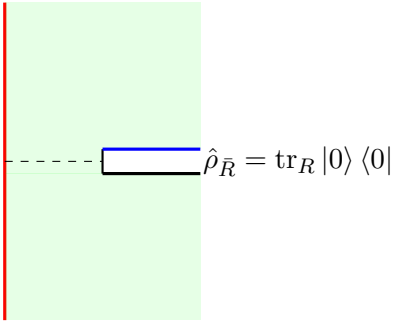
\begin{figure}[h]
    \centering
\begin{tikzpicture}[scale=0.65]
      \draw[-,very thick,red](0,-3) to (0,3.5);
       \draw[fill=green, draw=none, fill opacity = 0.1] (0,-3) to (4,-3) to (4,0) to (0,0);
       \draw[fill=green, draw=none, fill opacity = 0.1] (0,0.5) to (4,0.5) to (4,3.5) to (0,3.5);
       \draw[fill=green, draw=none, fill opacity = 0.1] (0,0.5) to (2,0.5) to (2,0) to (0,0);
       \draw[-,thick,black] (2,0.5) to (2,0);
       \draw[-,very thick,black] (2,0) to (4,0);
       \draw[-,very thick,blue] (2,0.5) to (4,0.5);
       \node at (6,0.25) {\textcolor{black}{$\hat{\rho}_{\bar{R}}=\tr_{R}\ket{0}\bra{0}$}};
       \draw[-,dashed,black] (0,0.25) to (2,0.25);
    \end{tikzpicture}
    \caption{\small  The Euclidean path integral preparing the ground state $\ket{0}$. The dashed interval denotes the subregion $R$. The resulting reduced density matrix operator $\hat{\rho}_{R}$ is obtained from $\hat{\rho}=\ket{0}\bra{0}$ by tracing over all possible field values $\phi_{R}$ in the subregion $R$. This operation effectively glued the black and blue slices in Fig.\ref{pic:preparerho} along the subregion $R$. If one is interested in the specific element of the resulting reduced density matrix operator $\bra{\phi_{\bar{R},1}}\hat{\rho}_{\bar{R}}\ket{\phi_{,\bar{R},2}}$ then one can specify the values of the fields $\phi^{*}_{\bar{R},2}(x)$ on the blue slice and $\phi_{\bar{R},1}(x)$ on the black slice.}
    \label{pic:reducedrho0}
\end{figure}

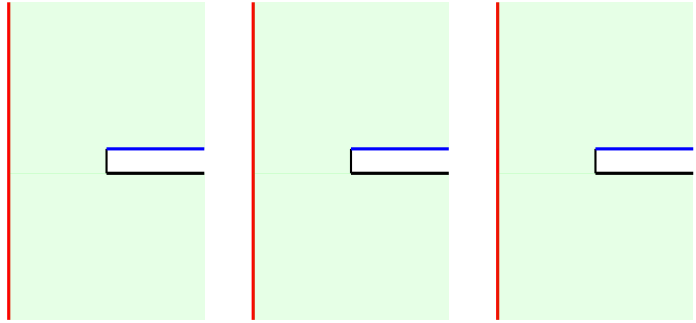
\begin{figure}[h]
    \centering
\begin{tikzpicture}[scale=0.65]
      \draw[-,very thick,red](0,-3) to (0,3.5);
       \draw[fill=green, draw=none, fill opacity = 0.1] (0,-3) to (4,-3) to (4,0) to (0,0);
       \draw[fill=green, draw=none, fill opacity = 0.1] (0,0.5) to (4,0.5) to (4,3.5) to (0,3.5);
       \draw[fill=green, draw=none, fill opacity = 0.1] (0,0.5) to (2,0.5) to (2,0) to (0,0);
       \draw[-,thick,black] (2,0.5) to (2,0);
       \draw[-,very thick,black] (2,0) to (4,0);
       \draw[-,very thick,blue] (2,0.5) to (4,0.5);

\draw[-,very thick,red](5,-3) to (5,3.5);
       \draw[fill=green, draw=none, fill opacity = 0.1] (5,-3) to (9,-3) to (9,0) to (5,0);
       \draw[fill=green, draw=none, fill opacity = 0.1] (5,0.5) to (9,0.5) to (9,3.5) to (5,3.5);
       \draw[fill=green, draw=none, fill opacity = 0.1] (5,0.5) to (7,0.5) to (7,0) to (5,0);
       \draw[-,thick,black] (7,0.5) to (7,0);
       \draw[-,very thick,black] (7,0) to (9,0);
       \draw[-,very thick,blue] (7,0.5) to (9,0.5);

       \draw[-,very thick,red](10,-3) to (10,3.5);
       \draw[fill=green, draw=none, fill opacity = 0.1] (10,-3) to (14,-3) to (14,0) to (10,0);
       \draw[fill=green, draw=none, fill opacity = 0.1] (10,0.5) to (14,0.5) to (14,3.5) to (10,3.5);
       \draw[fill=green, draw=none, fill opacity = 0.1] (10,0.5) to (12,0.5) to (12,0) to (10,0);
       \draw[-,thick,black] (12,0.5) to (12,0);
       \draw[-,very thick,black] (12,0) to (14,0);
       \draw[-,very thick,blue] (12,0.5) to (14,0.5);
    \end{tikzpicture}
    \caption{\small An example of the computation of $\tr \hat{\rho}_{\bar{R}}^{3}$ where one starts with three copies of the manifold appearing in Fig.\ref{pic:reducedrho0} and glue them by gluing the black slice of the $i$-th copy to the blue slice of the $i+1$-th copy and at the end glue the black slice of the last copy to the blue slice of the first copy. Then one obtains the replica manifold $\mathcal{M}^{(n)}$ (here $n=3$) and one just computes the Euclidean path integral of the field theory on this replica manifold.}
    \label{pic:trrhon}
\end{figure}

In the computation of the entanglement entropy between the subregion $R$ and its complement $\bar{R}$, we are interested in the reduced density matrix operator
\begin{equation}
\hat{\rho}_{\bar{R}}=\tr_{R}\ket{0}\bra{0}\,,
\end{equation}
which is obtained from Equ.~(\ref{eq:rho0}) by tracing out the degrees of freedom (i.e. possible field values) in the region $R$. This can again be prepared using a Euclidean path integral on the plane as shown in Fig.\ref{pic:reducedrho0}. With this density matrix equipped, we can compute the entanglement entropy using the replica trick
\begin{equation}
    S_{R}=-\tr\hat{\rho}_{\bar{R}}\log\hat{\rho}_{\bar{R}}=\lim_{n\rightarrow1}\frac{1}{1-n}\log\tr \hat{\rho}_{\bar{R}}^{n}\,,\label{eq:replicatrick}
\end{equation}
for which we have assumed that the reduced density matrix operator $\hat{\rho}_{\bar{R}}$ has been normalized to have a unit trace. One can compute $\tr \hat{\rho}_{\bar{R}}^{n}$ using Euclidean path integral as indicated in Fig.\ref{pic:trrhon} by cyclically gluing $n$ copies of the manifold in Fig.\ref{pic:reducedrho0} to a replica manifold $\mathcal{M}^{(n)}$ and computing the Euclidean path integral on $\mathcal{M}^{(n)}$. We note that the metric on the replica manifold contains a conical singularity at the coboundary of $R$ and $\bar{R}$ but the field is smooth on the replica manifold $\mathcal{M}^{(n)}$ (for a 2d CFT such a conical singularity can be regulated in various ways \cite{Sully:2020pza}). Let's denote the Euclidean path integral over smooth field configurations on $\mathcal{M}^{(n)}$ to be $Z_{\text{BCFT}}[\mathcal{M}^{(n)}]$. Let's denote the half Euclidean plane with the Cardy boundary by $\mathcal{M}$ then the trace of the n-th power of the normalized reduced density matrix operator is in fact given by
\begin{equation}
    \tr \hat{\rho}_{\bar{R}}^{n}=\frac{Z_{\text{BCFT}}[\mathcal{M}^{(n)}]}{Z_{\text{BCFT}}^{n}[\mathcal{M}]}\,.\label{eq:key}
\end{equation}
This is the key formula for later calculations and the proof of the Ryu-Takayanagi conjecture Equ.~(\ref{eq:RT}) in AdS/BCFT using gravitational path integral.

\subsubsection{Boundary Calculation}\label{sec:boundary}
Now let's apply Equ.~(\ref{eq:key}) to compute the subregion entanglement entropy in BCFT$_{2}$ for various situations. 

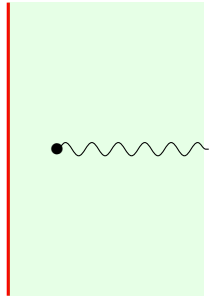
\begin{figure}[h]
    \centering
\begin{tikzpicture}[scale=0.65]
      \draw[-,very thick,red](0,-3) to (0,3);
       \draw[fill=green, draw=none, fill opacity = 0.1] (0,-3) to (4,-3) to (4,3) to (0,3);
       \draw[-,decorate,decoration=snake] (1,0) to (4.1,0);
       \node at (1,0) {\textcolor{black}{$\bullet$}};
    \end{tikzpicture}
    \caption{\small  half-plane with a twist operator inserted at $\tau=0,x=\ell$ (the black dot) which creates a branch cut for the field along $x\geq \ell$.}
    \label{pic:planetwist}
\end{figure}

To warm up, let's start with the simplest case, i.e. the case we discussed in Sec.\ref{sec:replicatrick} which is the boundary description of the set-up in Sec.\ref{sec:empty}. In this case, we have the replica manifold $\mathcal{M}^{(n)}$ as the $n$-branched half-plane with the branching point at the coboundary of $R$ and $\bar{R}$. Let's define the length of the subregion $R$ to be $\ell$ then the branching point is at $x=\ell$ in the geometry Equ.~(\ref{eq:plane}). The fields on the replica manifold $\mathcal{M}^{(n)}$ obey a $\mathbb{Z}_{n}$ symmetry which cyclically permutes the fields on different branches. Hence, one can consider the quotient case for which the background manifold becomes $\mathcal{M}=\mathcal{M}^{(n)}/\mathbb{Z}_{n}$ and the field becomes multi-valued at the original branch cut $x\geq \ell$. This is an orbifold BCFT$_{2}$ for which the original Euclidean path integral over smooth configurations on $\mathcal{M}^{(n)}$ is equal to the one-point function of a twist operator $\Phi_{n}(z,\bar{z})$ inserted at the branching point $x=\ell$ evaluated on $\mathcal{M}$ \cite{Calabrese:2009qy}. The twist operator creates a branch cut for the field along $\bar{R}$ i.e. $x\geq \ell$ and it is a primary operator with conformal weights
\begin{equation}
    h_{n}=\bar{h}_{n}=\frac{c}{24}(n-\frac{1}{n})\,,
\end{equation}
where $c$ is the central charge of the BCFT$_{2}$ (see Fig.\ref{pic:planetwist}). Hence we have
\begin{equation}
  S_{R}=\lim_{n\rightarrow1}\frac{1}{1-n}\ln\langle\Phi_{n}(z,\bar{z})\rangle_{\text{HP}} \,, \label{eq:von}
\end{equation}
which can be computed using the boundary operator expansion (BOE) \cite{Sully:2020pza,Geng:2021iyq} as
\begin{equation}
    S_{R}=\frac{c}{6}\ln(\frac{2\ell}{\epsilon})+\ln(g_{a}) \,,\label{eq:CFTEE0}
\end{equation}
where $\epsilon$ is a UV cutoff length scale, the boundary entropy is defined as
\begin{equation}
    \ln(g_{a})-\frac{c}{6}\ln(\epsilon)=\lim_{n\rightarrow1}\frac{1}{1-n}\ln(B_{\Phi_{n}1}^{a})\,,
    \label{eq:bdyentropy}
\end{equation}
and $B_{\Phi_{n}1}^{a}$ is the boundary operator expansion (BOE) coefficient of $\Phi_{n}(z,\bar{z})$ into the boundary identity operator when the boundary condition is a Cardy boundary condition denoted as $a$.

\begin{figure}[h]
    \centering
    \subfloat[]{
    \begin{tikzpicture}[scale=0.65]
        \draw[-,dashed,black] (-4,0) to (-1,0);
        \draw[-,dashed,black] (1,0) to (4,0);
        \draw[-,very thick,red] (-1,0) arc (-180:180:1);
        \draw[fill=green, draw=none, fill opacity = 0.1] (-4,-3) to (-4,0) to (-1,0) arc (-180:0:1) to (4,0) to (4,-3);
        \draw[fill=green, draw=none, fill opacity = 0.1] (-4,3) to (-4,0) to (-1,0) arc (180:0:1) to (4,0) to (4,3);
        \draw[-,thick,blue] (-0.707,0.707) to (-3,3);
        \draw[-,thick,blue] (0.707,0.707) to (3,3);
         \draw[-,very thick,red] (-1.5,1.5) to (-3,3);
        \draw[-,very thick,red] (1.5,1.5) to (3,3);
        \node at (-1.5,1.5) {\textcolor{black}{$\bullet$}};
        \node at (1.5,1.5) {\textcolor{black}{$\bullet$}};
        \node at (-1.4,2.5) {\textcolor{red}{$R_{I}$}};
        \node at (1.3,2.5) {\textcolor{red}{$R_{II}$}};
    \end{tikzpicture}}
    \hspace{1cm}
    \subfloat[]{
    \begin{tikzpicture}[scale=0.65]
        \draw[-,dashed,black] (-4,0) to (-1,0);
        \draw[-,dashed,black] (1,0) to (4,0);
        \draw[-,very thick,red] (-1,0) arc (-180:180:1);
        \draw[fill=green, draw=none, fill opacity = 0.1] (-4,-3) to (-4,0) to (-1,0) arc (-180:0:1) to (4,0) to (4,-3);
        \draw[fill=green, draw=none, fill opacity = 0.1] (-4,3) to (-4,0) to (-1,0) arc (180:0:1) to (4,0) to (4,3);
         \draw[-,decorate,decoration=snake] (-1.5,1.5) to (-3,3);
        \draw[-,decorate,decoration=snake] (1.5,1.5) to (3,3);
        \node at (-1.5,1.5) {\textcolor{black}{$\bullet$}};
        \node at (1.5,1.5) {\textcolor{black}{$\bullet$}};
    \end{tikzpicture}}
    \caption{\small \textbf{a)}The path integral preparation of the $\ket{\text{TFD}}$ state in the $w$-plane. The two solid black slices specifies the time-evolved state on which we show the subregions $R_{I}$ and $R_{II}$ in red. The twist operators are inserted at the black dots. \textbf{b)} The insertion of twist operators comes from the replica path integral and creates the branch cuts as indicated by the weaved lines.}
    \label{pic:TFD1insertion}
\end{figure}
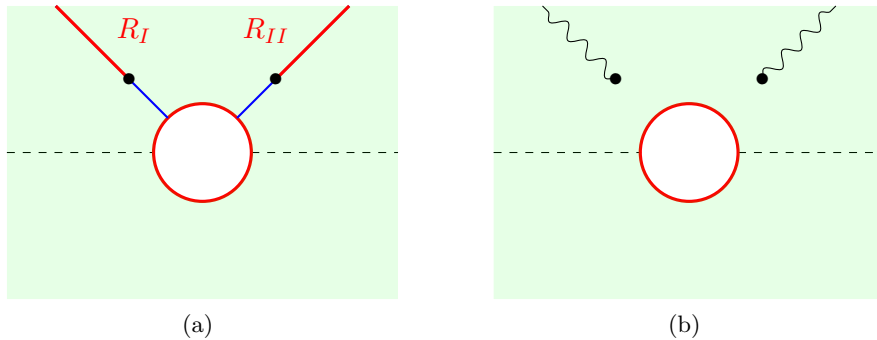

Now let's consider the thermal-field-double-state $\ket{\text{TFD}}$ of two BCFT$_{2}$'s. The BCFT$_{2}$'s all have one conformal boundary $a$ corresponding to a Cardy boundary condition.  We will label the two BCFT$_{2}$'s as BCFT$_{\text{L}}$ and BCFT$_{\text{R}}$ and this is the boundary description of the set-up in Sec.\ref{sec:BTZ1}. 
We can prepare the state $\ket{\text{TFD}}$ using an Euclidean path integral. To do this we firstly do a conformal transform from $z=x+i\tau$, for which the BCFT$_{\text{R}}$ part of the $\ket{\text{TFD}}$ state is supported at the slice $\tau=0$ and BCFT$_{L}$ part is at $\tau=\frac{\beta}{2}$, to
\begin{equation}
    w=e^{\frac{2\pi}{\beta}z}\,,\quad \bar{w}=e^{\frac{2\pi}{\beta}\bar{z}}\,,\label{eq:conformaltransform}
\end{equation}
where $\beta$ is the inverse temperature. The path integral is over the region from the $\tau=0$ slice to the $\tau=\frac{\beta}{2}$ slice clockwisely and in the $w$-plane can be represented as in Fig.\ref{pic:TFD1}. We want to compute the entanglement entropy of the subregion $R=R_{I}\cup R_{II}$ as indicated in the bath of Fig.\ref{pic:penroseoriginal}. By the discussion of the previous paragraph, this corresponds to the insertion of twist operators on the $w$-plane \cite{Sully:2020pza,Geng:2021iyq} at $\partial R_{I}$ and $\partial R_{II}$ (see Fig.\ref{pic:TFD1insertion}). We denote the two insertion points as $w_{L}=-e^{\ell\frac{2\pi}{\beta}-i\tau \frac{2\pi}{\beta}}$ and $w_{R}=e^{\ell\frac{2\pi}{\beta}+i\tau\frac{2\pi}{\beta}}$ where the Euclidean time $\tau$ is related to the Lorentzian time $t$ as $\tau=it$. Hence we have
\begin{equation}
  S_{R}=\lim_{n\rightarrow1}\frac{1}{1-n}\ln\langle\Phi_{n,L}(z,\bar{z})\Phi_{n,R}(z,\bar{z})\rangle_{\text{TFD}} \,, \label{eq:von2}
\end{equation}
for which the two-point function can be computed on the $w$-plane as the ground state two-point function
\begin{equation}
    \begin{split}
        \langle\Phi_{n,L}(z,\bar{z})\Phi_{n,R}(z,\bar{z})\rangle_{\text{TFD}}=\abs{\frac{dw}{dz}}_{w=w_{L}}^{2h_{n}}\abs{\frac{dw}{dz}}_{w=w_{R}}^{2h_{n}}\langle\Phi_{n}(w_{L},\bar{w}_{L})\Phi_{n}(w_{R},\bar{w}_{R})\rangle\,.
    \end{split}
\end{equation}
As a result, for holographic BCFT$_{2}$'s, we have \cite{Sully:2020pza,Geng:2021iyq}
\begin{equation}
    S_{R}=\min\Bigg[\frac{c}{3}\ln(\frac{\beta e^{\frac{2\pi\ell}{\beta}}\cosh \frac{2\pi t}{\beta} }{\epsilon\pi})\,,\frac{c}{3}\ln(\frac{\beta(e^{\frac{4\pi\ell}{\beta}}-1)}{2\epsilon\pi})+2\ln(g_{a})\Bigg] \,,\label{eq:CFTEE1}
\end{equation}
where the first candidate result under minimization is the bulk-channel result and the second is the boundary-channel result. In each channel we only have to consider the contribution from the identity operator in the operator product expansion which is a property of a holographic BCFT.

\begin{figure}[h]
    \centering
    \begin{tikzpicture}[scale=0.65]
        \draw[-,very thick,black] (-4,0) to (-1,0);
        \draw[-,very thick,black] (1,0) to (4,0);
        \draw[-,very thick,red] (-1,0) arc (-180:0:1);
        \draw[fill=green, draw=none, fill opacity = 0.1] (-4,-3) to (-4,0) to (-1,0) arc (-180:0:1) to (4,0) to (4,-3);
        \node at (2.5,0.5) {\textcolor{black}{$\text{BCFT}_{\text{R}}$}};
         \node at (-2.5,0.5) {\textcolor{black}{$\text{BCFT}_{\text{L}}$}};
    \end{tikzpicture}
    \caption{\small The path integral preparation of the state $\ket{\text{TFD}}$ in the $w$-plane. The black slices specify the support of the state $\ket{\text{TFD}}$.}
    \label{pic:TFD1}
\end{figure}
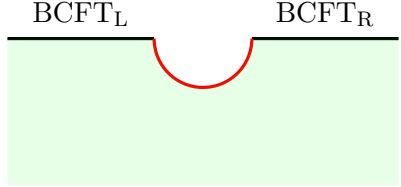

\begin{figure}[h]
    \centering
    \begin{tikzpicture}[scale=0.65]
        \draw[-,very thick,black] (-4,0) to (-1,0);
        \draw[-,very thick,black] (1,0) to (4,0);
        \draw[-,very thick,red] (-1,0) arc (-180:0:1);
        \draw[-,very thick,orange] (-4,0) arc (-180:0:4);
        \draw[fill=green, draw=none, fill opacity = 0.1] (-4,0) to (-1,0) arc (-180:0:1) to (4,0) arc (0:-180:4);
        \node at (2.5,0.5) {\textcolor{black}{$\text{BCFT}_{\text{R}}$}};
         \node at (-2.5,0.5) {\textcolor{black}{$\text{BCFT}_{\text{L}}$}};
    \end{tikzpicture}
    \caption{\small The path integral preparation of the state $\ket{\text{TFD}}$ in the $w$-plane. The black slices specify the support of the state $\ket{\text{TFD}}$.}
    \label{pic:TFD2}
\end{figure}
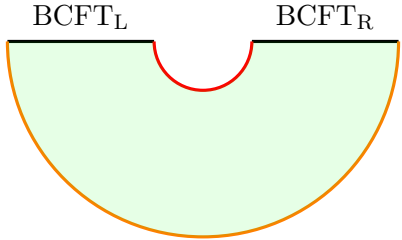

The last case we will consider is the thermal-field-double-state $\ket{\text{TFD}}$ of two BCFT$_{2}$'s where the two BCFT$_{2}$'s both involve two conformal boundaries corresponding to Cardy boundary conditions $a$ and $b$. This is the boundary description of the set-up in Sec.\ref{sec:BTZ2}. Let the distance between the two boundaries be $L$. The state $\ket{\text{TFD}}$ can be prepared in the same way as in the one boundary case by conformally transforming from $z=x+i\tau$ to $w=e^{\frac{2\pi}{\beta}z}$ (see Fig.\ref{pic:TFD2}). The entanglement entropy calculation is also the same as in the last paragraph. Nevertheless, now due to the appearance of another boundary the result is different \cite{Geng:2021iyq}. For a holographic BCFT we have
\begin{equation}
        S_{\mathcal{A}}=\min\Bigg[\frac{c}{3}\ln(\frac{\beta(e^{\frac{2\pi \ell}{\beta}}-1}{2\epsilon\pi})+2\ln(g_{a})\,,\frac{c}{3}\ln(\frac{2 e^{\frac{4\pi\ell}{\beta}}\cosh \frac{2\pi t}{\beta} }{\epsilon\pi})\,,\frac{c}{3}\ln(\frac{\beta(e^{\frac{2\pi L}{\beta}}-e^{\frac{2\pi}{\beta}(2\ell-L)})}{2\epsilon\pi})+2\ln(g_{b})\Bigg] \,,\label{eq:CFTEE2}
\end{equation}
where the new candidate under the minimization compared with Equ.~(\ref{eq:CFTEE1}) corresponds to the boundary channel operator product expansion associated with the second boundary.

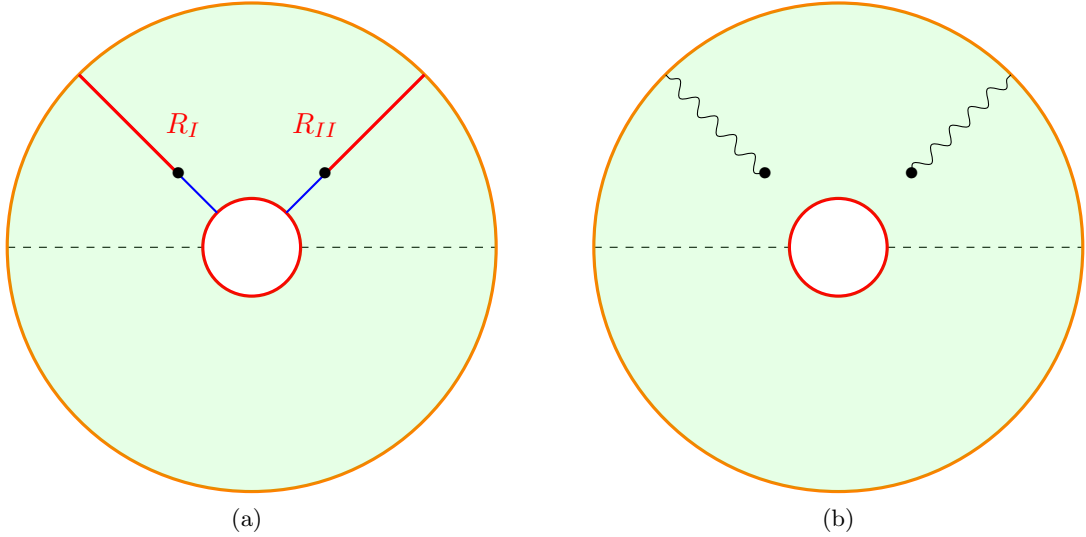
\begin{figure}[h]
    \centering
    \subfloat[]{
    \begin{tikzpicture}[scale=0.65]
        \draw[-,dashed,black] (-5,0) to (-1,0);
        \draw[-,dashed,black] (1,0) to (5,0);
        \draw[-,very thick,red] (-1,0) arc (-180:180:1);
        \draw[-,very thick,orange] (-5,0) arc (-180:180:5);
        \draw[fill=green, draw=none, fill opacity = 0.1] (-5,0) to (-1,0) arc (-180:0:1) to (5,0) arc (0:-180:5);
        \draw[fill=green, draw=none, fill opacity = 0.1] (-5,0) to (-1,0) arc (180:0:1) to (5,0) arc (0:180:5);
        \draw[-,thick,blue] (-0.707,0.707) to (-3.53,3.53);
        \draw[-,thick,blue] (0.707,0.707) to (3.53,3.53);
         \draw[-,very thick,red] (-1.5,1.5) to (-3.53,3.53);
        \draw[-,very thick,red] (1.5,1.5) to (3.53,3.53);
        \node at (-1.5,1.5) {\textcolor{black}{$\bullet$}};
        \node at (1.5,1.5) {\textcolor{black}{$\bullet$}};
        \node at (-1.4,2.5) {\textcolor{red}{$R_{I}$}};
        \node at (1.3,2.5) {\textcolor{red}{$R_{II}$}};
    \end{tikzpicture}}
    \hspace{1cm}
    \subfloat[]{
     \begin{tikzpicture}[scale=0.65]
  \draw[-,dashed,black] (-5,0) to (-1,0);
        \draw[-,dashed,black] (1,0) to (5,0);
        \draw[-,very thick,red] (-1,0) arc (-180:180:1);
        \draw[-,very thick,orange] (-5,0) arc (-180:180:5);
        \draw[fill=green, draw=none, fill opacity = 0.1] (-5,0) to (-1,0) arc (-180:0:1) to (5,0) arc (0:-180:5);
        \draw[fill=green, draw=none, fill opacity = 0.1] (-5,0) to (-1,0) arc (180:0:1) to (5,0) arc (0:180:5);
        \node at (-1.5,1.5) {\textcolor{black}{$\bullet$}};
        \node at (1.5,1.5) {\textcolor{black}{$\bullet$}};
    \draw[-,decorate,decoration=snake] (-1.5,1.5) to (-3.53,3.53);
        \draw[-,decorate,decoration=snake] (1.5,1.5) to (3.53,3.53);
        \end{tikzpicture}
    }
    \caption{\small \textbf{a)}The path integral preparation of the $\ket{\text{TFD}}$ state in the $w$-plane. The two solid black slices specifies the time-evolved state on which we show the subregions $R_{I}$ and $R_{II}$ in red. The twist operators are inserted at the black dots. \textbf{b)} The insertion of twist operators comes from the replica path integral and creates the branch cuts as indicated by the weaved lines.}
    \label{pic:TFD2insertion}
\end{figure}

\subsubsection{Bulk Calculation}\label{sec:bulk}
We can also compute the subregion entanglement entropies in Sec.\ref{sec:boundary} in the corresponding bulk descriptions using the Ryu-Takayanagi conjecture Equ.~(\ref{eq:RT}). These computations were performed in detail in \cite{Sully:2020pza,Geng:2021iyq} with the results exactly matching Equ.~(\ref{eq:CFTEE0}), Equ.~(\ref{eq:CFTEE1}) and Equ.~(\ref{eq:CFTEE2}). Here we review relevant geometric aspects for our later discussion of the replica wormholes.

In the simplest case we considered, i.e. the ground state of a BCFT$_{2}$ with single Cardy boundary $a$, the bulk description is as in Fig.\ref{pic:emptyads3}. The entanglement entropy $S_{R}$ can be computed by looking for a minimal area surface connecting $\partial R$ to the Karch-Randall brane in the bulk description. One also has to minimize the area over the ending point on the brane which as we discussed corresponds to implementing the quantum extremal surface prescription in the intermediate description. The resulting configuration is shown in Fig.\ref{pic:emptyads3}. The nice feature of this example is that there exists entanglement island even without a black hole and its very existence is guaranteed by the nonzero value of the entanglement entropy $S_{R}$. Hence we can see that entanglement island is an essential element in holography.

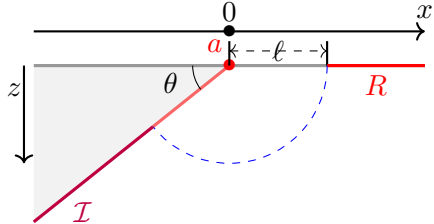
\begin{figure}[h]
\begin{centering}
\begin{tikzpicture}[scale=0.65]
\draw[-,very thick,black!40] (-4,0) to (0,0);
\draw[-,very thick,black!40] (0,0) to (4,0);
\draw[-,very thick,red!60] (0,0) to (-1.56,-1.25);
\draw[fill=gray, draw=none, fill opacity = 0.1] (0,0) to (-4,-3.2) to (-4,0) to (0,0);
\draw[-] (-0.75,0) arc (180:217.5:0.8);
\node at (-0.3,0.4) {\textcolor{red}{$a$}};
\node at (-1.2,-0.35) {$\theta$};
\node at (0,0.7) {$\bullet$};
\node at (0,1.1) {$0$};
\node at (0,0) {\textcolor{red}{$\bullet$}};
\draw[->,thick,color=black] (-4,0.7) to (4,0.7);
\node at (4,1.1)
{\textcolor{black}{$x$}};
\draw[->,thick,color=black] (-4.2,0) to (-4.2,-2);
\node at (-4.4,-0.5)
{\textcolor{black}{$z$}};
\draw[-,thick,black] (0,0) to (0,0.5);
\draw[-,thick,black] (2,0) to (2,0.5);
\draw[<->,dashed, black] (0,0.3) to (2,0.3);
\node at (1,0.3) {\textcolor{black}{\textbf{$\ell$}}};
\draw[-,very thick,red] (2,0) to (4,0);
\node at (3,-0.4) {\textcolor{red}{$R$}};
\draw[-,dashed,blue] (2,0) arc (0:-141.3:2); 
\draw[-,very thick,purple!100] (-1.56,-1.25) to (-4,-3.2);
\node at (-3,-3) {\textcolor{purple}{$\mathcal{I}$}};
\end{tikzpicture}
\caption{\small The computation of $S_{R}$ using RT conjecture for the ground state. The dashed blue curve is the RT surface and the purple region $\mathcal{I}$ on the Karch-Randall brane is the entanglement island of the bath region $R$ in the intermediate description.}
\label{pic:emptyads3RT}
\end{centering}
\end{figure}

\begin{figure}[h]
     \centering
     \begin{tikzpicture}[scale=0.65]
         \draw[fill=green, draw=none, fill opacity = 0.1] (-6,0) to (-2,2.5) to (10,2.5) to (6,0);
         \draw[-,very thick,red] (3.6,1.5) arc (0:-180:2 and 1);
         \draw[dashed,very thick,red] (3.6,1.5) arc (0:180:2 and 0.7);
         \draw[-,very thick,red] (3.6,1.5) arc (20:160:2.128);
         \draw[fill=red, draw=none, fill opacity = 0.2] (3.6,1.5) arc (0:-180:2 and 1) arc (160:20:2.128);
         \draw[->,thick,black] (-4,1.5) to (-4,4);
         \node at (-4.4,4) {\textcolor{black}{$z$}};
         \draw[dashed,thick,black] (3.6,1.5) to (8.4,1.5);
         \draw[dashed,thick,black] (-0.4,1.5) to (-3.6,1.5);
     \end{tikzpicture}
     \caption{\small The configuration of the Karch-Randall braneworld Equ.~(\ref{eq:KRBTZ2}) in the geometry Equ.~(\ref{eq:metric2}). The brane is the red spherical cap and the interior of the cap is removed from the bulk geometry. The $\ket{\text{TFD}}$ is supported at the zero time $\tau=0$ slice which is the dashed black line.}
     \label{pic:EAdS3BTZ1}
 \end{figure}
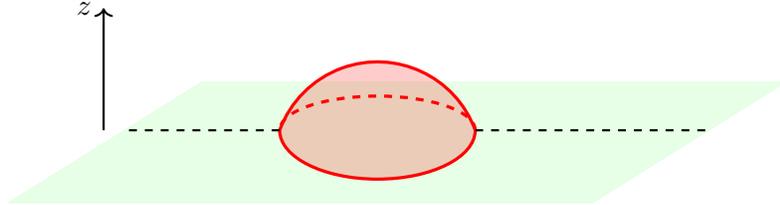

In the case of two BCFT$_{2}$'s with single Cardy boundary in the thermal-field-double-state $\ket{\text{TFD}}$, the bulk description contains a BTZ black hole and a Karch-Randall brane as indicated in Fig.\ref{pic:AdS3BTZ1}. The computation of the subregion entanglement entropy using the Ryu-Takayanagi conjecture Equ.~(\ref{eq:RT}) is essentially the same as in Fig.\ref{pic:KRrealization}. Nevertheless, as we have seen in Sec.\ref{sec:boundary}, it is important to perform the conformal transform Equ.~(\ref{eq:conformaltransform}) for the replica computation of the entanglement entropy in the boundary description. For the sake of later convenience, here we discuss the corresponding transformation in the bulk description. In the bulk, we start from the Euclidean BTZ metric
\begin{equation}
     ds^2=\frac{h(z)}{z^2}d\tau^2+\frac{dz^2}{h(z)z^2}+\frac{dx^2}{z^2} \,,\qquad h(z)=1-\frac{z^2}{z_{h}^2} \,,  \label{eq:EBTZ}
\end{equation}
for which $\tau$ is periodic with a period $\beta=2\pi z_{h}$ as the inverse temperature. A Karch-Randall brane is embedded as
\begin{equation}
    x(z)=- z_h \arcsinh{\left(\frac{z\,T}{z_{h}\sqrt{1-T^2}}\right)} \,.\label{eq:KRBTZ}
\end{equation}
The following bulk coordinate transformation corresponds to the boundary conformal transform Equ.~(\ref{eq:conformaltransform})
\begin{equation}
    \begin{split}
        z'=\frac{z}{z_{h}}e^{\frac{z}{x_{h}}}\,,\quad w=e^{\frac{x+i\tau}{z_{h}}}\sqrt{h(z)}\,,\label{eq:bulktransform}
    \end{split}
\end{equation}
which transforms the metric Equ.~(\ref{eq:EBTZ}) to
\begin{equation}
    ds^2=\frac{dz'^2+dwd\bar{w}}{z'^2}\,.\label{eq:metric2}
\end{equation}
Near the conformal boundary $z'=\epsilon\rightarrow0$, the transformation Equ.~(\ref{eq:bulktransform}) is exactly the boundary conformal transform Equ.~(\ref{eq:conformaltransform}). Moreover, the fact that field theory in the $w$-plane is in the ground state exactly matches the result Equ.~(\ref{eq:metric2}) that the bulk metric is now empty AdS$_{3}$. Under the above coordinate transformation the Karch-Randall brane Equ.~(\ref{eq:KRBTZ}) is now embedded in Equ.~(\ref{eq:metric2}) as
\begin{equation}
    \frac{1}{1-T^2}=w\bar{w}+(z'+\frac{T}{\sqrt{1-T^2}})^2\,,\label{eq:KRBTZ2}
\end{equation}
which is a spherical cap (see Fig.\ref{pic:EAdS3BTZ1}) and near the conformal boundary $z'=\epsilon\rightarrow0$ it is exactly $w\bar{w}=1$ which matches with Fig.\ref{pic:TFD1insertion}.

The case where the boundary description involves two BCFT$_{2}$'s each with two Cardy boundaries $a$ and $b$ in $\ket{\text{TFD}}$ state is pretty much the same as the one boundary case except now the bulk description contains two branes both of which are spherical caps in the geometry Equ.~(\ref{eq:metric2}). Moreover, the Ryu-Takayanagi surface can now end on either of the two branes which reproduces exactly Equ.~(\ref{eq:CFTEE2}) \cite{Geng:2021iyq}.

 \section{Replica Wormholes in the Karch-Randall Braneworld}\label{sec:rw}
In the previous section we see that in the case where the bulk description is asymptotically AdS$_{3}$ with Karch-Randall branes and the associated boundary description is a BCFT$_{2}$, the computations of the subregion entanglement entropy $S_{R}$ using the BCFT$_{2}$ techniques on the boundary and the Ryu-Takayanagi conjecture Equ.~(\ref{eq:RT}) in the bulk give exactly the same answers in various situations. In this calculation, the entanglement island naturally emerges on the Karch-Randall brane.  

In this section, we give a proof of the Ryu-Takayanagi conjecture Equ.~(\ref{eq:RT}) in the above context following the work \cite{Lewkowycz:2013nqa,Dong:2016hjy,Dong:2017xht}. Our strategy is to translate the boundary replica path integral Equ.~(\ref{eq:key}) into a bulk gravitational path integral using the holographic dictionary then evaluate it using the saddle point approximation to first order in $(n-1)$-expansion and finally derive a formula for $S_{R}$ following Equ.~(\ref{eq:replicatrick}).

\subsection{Replica Wormhole and the Ryu-Takayanagi Conjecture}\label{sec:proof}

\begin{figure}[h]
    \centering
    \includegraphics[scale=0.2]{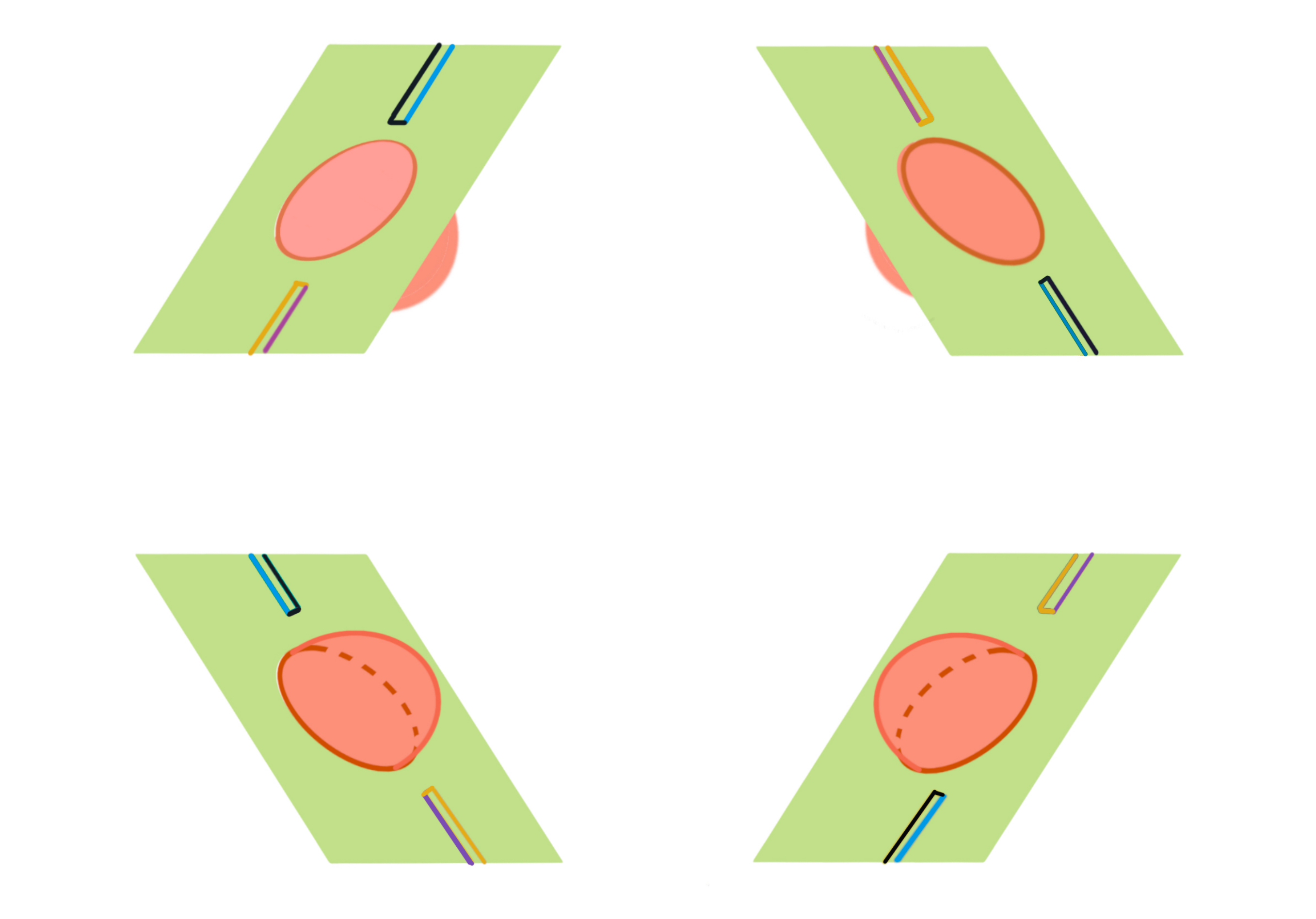}
    \caption{\small The disconnected phase of the Karch-Randall brane in the replica path integral for $n=4$. The green surfaces are asymptotic boundaries of the bulk and the red spherical caps are the branes. The bulk region inside the spherical caps is cut off. In the boundary, the black slice in the $i$-th component should be glued to the blue slice in the $(i+1)$-th component and the orange slice of the $i$-th component should be glued to the purple slice of the $(i+1)$-th boundary. For the sake of convenience we in fact drew the case for the state $\ket{\text{TFD}}$ without any time evolution. The time-evolved case is easy to visualize according to Fig.\ref{pic:TFD1insertion} but hard to draw here.}
    \label{pic:disconnected}
\end{figure}

\begin{figure}[h]
    \centering
    \includegraphics[scale=0.25]{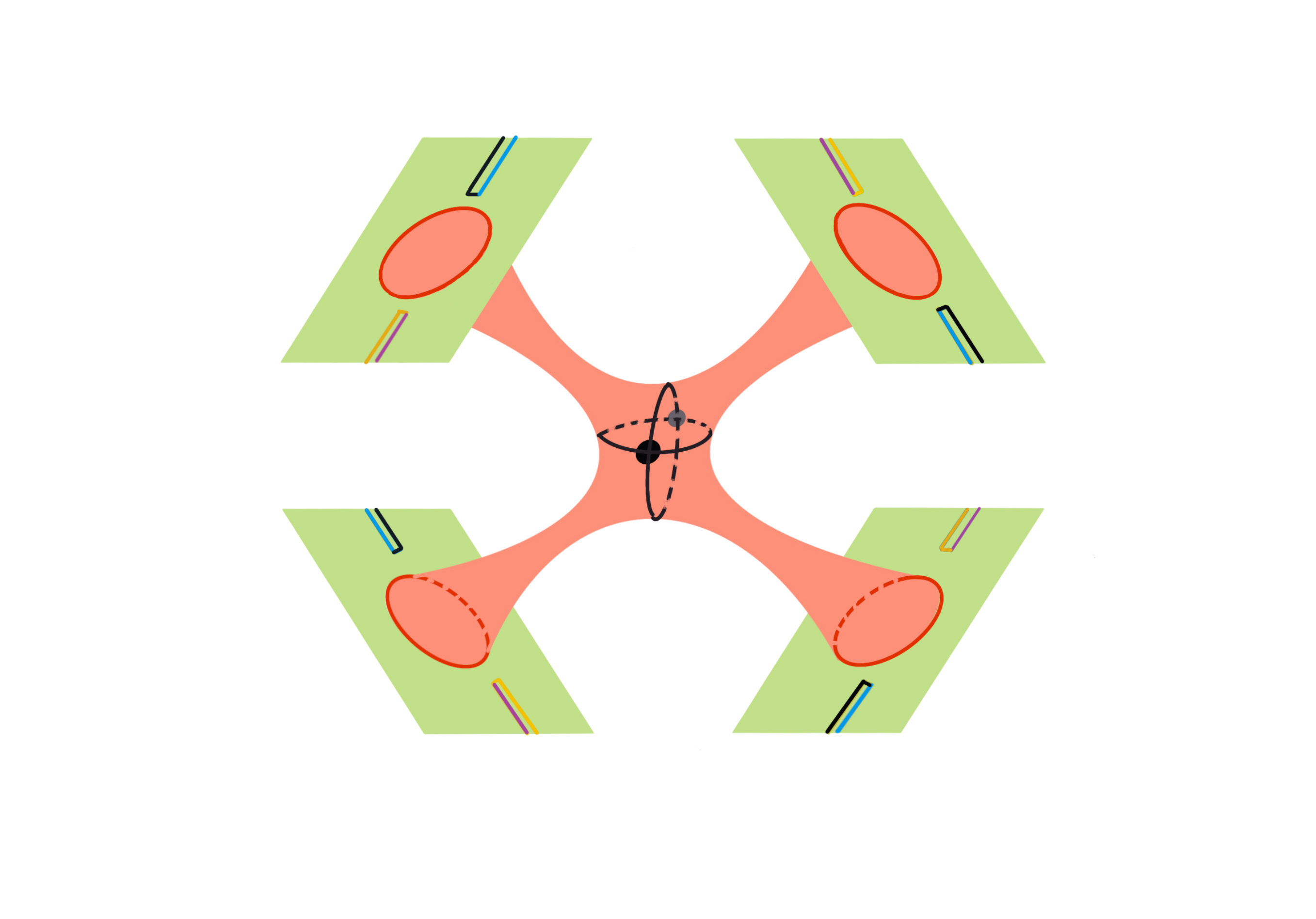}
    \caption{\small The connected phase of the Karch-Randall brane in the replica path integral for $n=4$. The green surfaces are asymptotic boundaries of the bulk and the red surface is the brane. The bulk region inside the brane is cut off. In the bulk, the fix point locus of the $\mathbb{Z}_{n}$ symmetry extends to the brane. The two black dots indicate the fixed points of $\mathbb{Z}_{n}$ symmetry on the brane. In the boundary, the black slice in the $i$-th component should be glued to the blue slice in the $(i+1)$-th component and the orange slice of the $i$-th component should be glued to the purple slice of the $(i+1)$-th boundary. For the sake of convenience we in fact drew the case for the state $\ket{\text{TFD}}$ without any time evolution. The time-evolved case is easy to visualize according to Fig.\ref{pic:TFD1insertion} but hard to draw here.}
    \label{pic:connected}
\end{figure}

We want to compute the subregion entanglement entropy $S_{\text{R}}$ of the thermal-field-double-state in the case as shown in Fig.\ref{pic:TFD1insertion}. Using the holographic dictionary of the Karch-Randall braneworld we can map the replicated boundary partition function into a Euclidean gravitational path integral
\begin{equation}
    Z_{\text{BCFT}}[\mathcal{M}^{(n)}]=\int_{\partial\mathcal{M}^{(n)}_{d+1}=\mathcal{M}^{(n)}} D[g] e^{-S}=Z_{\text{grav}}[\mathcal{M}^{(n)}]\,,
\end{equation}
where the gravitational action $S$ is given by Equ.~(\ref{eq:action1}), the bulk manifold $\mathcal{M}^{(n)}_{d+1}$ should have the conformal boundary as the replica manifold $\mathcal{M}^{(n)}$ together with Karch-Randall branes $\mathcal{M}^{(n)}_{d}$ whose asymptotic boundary $\partial\mathcal{M}_{d}$ is the same as $\partial\mathcal{M}^{(n)}$ and we are integrating over smooth metric configurations, i.e. summing over smooth bulk geometries, as the field configurations on the boundary manifold $\mathcal{M}^{(n)}$ are smooth. The gravitational path integral can be computed using the saddle point approximation as the bulk gravitational theory is classical, i.e. $G_{N}$ is small. As a result, one has
\begin{equation}
     Z_{\text{BCFT}}[\mathcal{M}^{(n)}]=Z_{\text{grav}}[\mathcal{M}^{(n)}]=e^{-S[g_{0}]}\,,
\end{equation}
where the metric $g_{0}$ is smooth and satisfies the bulk Einstein's equation
\begin{equation}
R_{\mu\nu}-\frac{1}{2}g_{\mu\nu} R=-\Lambda g_{\mu\nu}=\frac{d(d-1)}{2}g_{\mu\nu}\,,\label{eq:Einstein}
\end{equation}
and the brane satisfies the brane embedding equation Equ.~(\ref{eq:braneemb}). Since the boundary manifold $\mathcal{M}^{(n)}$ has the replica symmetry $\mathbb{Z}_{n}$, one expects this symmetry to extend into the bulk. So the bulk metric $g_{0}$ has a $\mathbb{Z}_{n}$ symmetry, the boundary restriction of which is the boundary $\mathbb{Z}_{n}$ symmetry. Let's denote the bulk manifold with metric $g_{0}$ as $\mathcal{M}^{n}_{d+1}$. When we are computing the on-shell action $S[g_{0}]$ we can consider instead $n$ copies of the quotient manifold
\begin{equation}
\tilde{\mathcal{M}}_{d+1}=\mathcal{M}^{n}_{d+1}\slash\mathbb{Z}_{n}\,,\label{eq:quotient}
\end{equation}
which nevertheless contains a conical singularity corresponding to the locus of the $\mathbb{Z}_{n}$ fixed points in $\mathcal{M}^{n}_{d+1}$. We want to evaluate $S[g_{0}]$ to first order in $(n-1)$. Let's denote the metric of the quotient manifold Equ.~(\ref{eq:quotient}) as $\tilde{g}_{0}$. The conical singularity is a codimension-two bulk minimal surface $\gamma$ homologous to $R$ \cite{Lewkowycz:2013nqa}. When there are multiple such surfaces one should take the one for which the on-shell action is smallest which is due to the saddle point approximation we are using. In Einstein's gravity this conical singularity can be taken as a cosmic brane with tension
\begin{equation}
    \mathcal{T}_{n}=\frac{1}{4G_{d+1}}(1-\frac{1}{n})\,.
\end{equation}
Hence we have 
\begin{equation}
S[g_{0}]=nS[\tilde{g}_{0}]+\frac{A(\gamma)}{4G_{d+1}}\frac{n-1}{n}\,,
\end{equation}
where $A(\gamma)$ is the area of the codimension-two minimal area surface $\gamma$. Here we notice that $\tilde{g_{0}}$ depends on $n$ and when $n=0$ $\tilde{g_{0}}$ is just the Euclidean metric Equ.~(\ref{eq:EBTZ}). Since the Euclidean metric Equ.~(\ref{eq:EBTZ}) solves the bulk Einstein's equation, it is a stationary point of the action $S[g]$ so any variation around it wouldn't contribute to the action $S[g]$ to the first order of the variational parameter. In our case, the variational parameter is $(n-1)$ so to first order in $(n-1)$ we have 
\begin{equation}
    S[\tilde{g_{0}}]=S[g_{\text{EBTZ}}]\,.
\end{equation}
Therefore, we have
\begin{equation}
S[g_{0}]=nS[g_{\text{EBTZ}}]+\Big(\frac{A(\gamma)}{4G_{d+1}}\frac{n-1}{n}\Big)+\mathcal{O}((1-n)^2)\,,
\end{equation}
where the minimal area surface $\gamma$ now lives in the bulk geometry Equ.~(\ref{eq:EBTZ}) and is homologous to $R$. As a result, we have
\begin{equation}
\begin{split}
    S_{R}&=-\tr\hat{\rho}_{\bar{R}}\log\hat{\rho}_{\bar{R}}=\lim_{n\rightarrow1}\frac{1}{1-n}\log\tr \hat{\rho}_{\bar{R}}^{n}\,,\\&=\lim_{n\rightarrow1}\frac{1}{1-n}\frac{Z_{\text{grav}}[\mathcal{M}^{(n)}]}{Z_{\text{grav}}[\mathcal{M}^{(0)}]^{n}}\,,\\&=\lim_{n\rightarrow1}\frac{1}{1-n}\frac{e^{-nS[g_{\text{EBTZ}}]+\frac{A(\gamma)}{4G_{d+1}}\frac{1-n}{n}]}}{e^{-nS[g_{\text{EBTZ}}]}}\,,\\&=\frac{A(\gamma)}{4G_{d+1}}\,.
    \end{split}
\end{equation}
Nonetheless, in the above discussion we haven't carefully treat the Karch-Randall brane. We notice that there are two possible phases of the Karch-Randall brane in the manifold $\mathcal{M}_{d+1}^{(n)}$. The first phase has $n$-disconnected components (see Fig.\ref{pic:disconnected}) and the second phase has a single component connecting the Cardy boundaries of the boundary replica manifold $\mathcal{M}^{(n)}$ (see Fig.\ref{pic:connected}). In the former case the conical singularity $\gamma$ stays in the bulk connecting the two boundaries of $\partial R$ (the orange curve in Fig.\ref{pic:KRrealization}) and in the later case the conical singularity $\gamma$ extends to the brane as there also exist $\mathbb{Z}_{n}$ fixed points on the brane so $\gamma$ has two disconnected components connecting the two boundaries of $\partial R$ to the brane (the green curve in Fig.\ref{pic:KRrealization}). Thus, we in fact have
\begin{equation}
    S(R)=\min (\frac{A_{c}}{4G_{d+1}},\frac{2A_{dc}}{4G_{d+1}})\,,\label{eq:RTderiation}
\end{equation}
which is exactly the Ryu-Takayanagi conjecture Equ.~(\ref{eq:RT}). Moreover, we notice that when we are looking for $\gamma_{dc}$ we should also minimize its area over its possible ending points on the brane which is again according to the saddle point approximation we are using.

As a summary, we see that we can in fact prove the Ryu-Takayanagi conjecture Equ.~(\ref{eq:RT}). In the proof we see that the emergence of the disconnected Ryu-Takayanagi surface is due to the connected phase of the Karch-Randall brane in the replica path integral. When the connected phase dominates, the fixed point locus of the bulk $\mathbb{Z}_{n}$ symmetry extends to the brane which explains why the Ryu-Takayanagi surface could end on the brane. In the intermediate description of the Karch-Randall braneworld this is exactly the formation of the replica wormhole (see Fig.\ref{pic:connected}) and as a result the entanglement island as explained in \cite{Almheiri:2019qdq}.

\subsection{The Existence of the Replica Wormhole}\label{sec:exsitence}
In Sec.\ref{sec:proof} we see that the possible existence of the replica wormhole proves the emergence of the disconnected Ryu-Takayanagi surface and hence the formation of entanglement island on the brane. Nevertheless, since we are using the saddle-point approximation to compute the replica gravitational path integral, we also have to show that the connected phase of the Karch-Randall brane is consistent with the bulk Einstein's equation.

Let's fix metric to the Fefferman-Graham gauge \cite{fefferman2008ambient} by considering the following form of the bulk metric
\begin{equation}
    ds^2=dr^2+\bar{g}_{ab}(x,r) dx^{a} dx^{b}\,,\quad \text{where $a,b=1,2$}\,,
\end{equation}
for which the asymptotic boundary is $r\rightarrow\infty$ where $\bar{g}_{ab}$ decays as\footnote{The other part of the asymptotic boundary $r\rightarrow-\infty$ is cut off by the Karch-Randall brane.}
\begin{equation}
  \bar{g}_{ab}(x,r)\sim e^{-2r}\,, 
\end{equation}
and the Karch-Randall brane is located at a constant-$r$ slice $r=r_{B}$.\footnote{More precisely, we in fact expect in our case the the fall of behaviour of $\bar{g}_{ab}$ goes as
\begin{equation}
    \bar{g}_{ab}(x,r)\sim \cosh^2 r g_{ab}^{\text{AdS$_{d}$}}+\mathcal{O}(e^{-4r})\,,\quad\text{as $r\rightarrow\infty$}\,,
\end{equation}
where in the current case we have $d=2$.} Now using the results in the Sec.\ref{sec:app}, the brane embedding equation Equ.~(\ref{eq:braneemb}) reads
\begin{equation}
    \frac{1}{2}\partial_{r} \bar{g}_{ab}(x,r_{B})-\frac{1}{2}\text{Tr}(\bar{g}^{-1}\partial_{r}\bar{g})(x,r_{B}) \bar{g}_{ab}(x,r_{B})=T \bar{g}_{ab}(x,r_{B})\,,\label{eq:braneeq}
\end{equation}
which reduces to 
\begin{equation}
    -\frac{1}{2}\text{Tr}(\bar{g}^{-1}\partial_{r}\bar{g})(x,r_{B})=2T\,\label{eq:bcbrane}
\end{equation}
by taking the trace. Then combining Equ.~(\ref{eq:bcbrane}) with Equ.~(\ref{eq:braneeq}) we get
\begin{equation}
    \partial_{r}\bar{g}_{ab}(x,r_{B})=-2T \bar{g}_{ab}(x,r_{B})\,. \label{eq:bcbrane1}
\end{equation}
The bulk Einstein's equation Equ.~(\ref{eq:Einstein}) reads ($d=2$)
\begin{equation}
    R_{\mu\nu}-\frac{1}{2} g_{\mu\nu} R=g_{\mu\nu}\,.
\end{equation}
We are interested in the constraint it imposes on the geometry of the Karch-Randall brane, so we will evaluate the Einstein's equation at $r=r_{B}$. Using the results in Sec.\ref{sec:app}, we have
\begin{equation}
    \begin{split}
\partial_{r}\Big(\text{Tr}(\bar{g}^{-1}\partial_{r}\bar{g})\Big)(x,r_{B})+4T^{2}=4\,,\\
\bar{R}_{ab}(x,r_{B})=-2\bar{g}_{ab}(x,r_{B})+\frac{1}{2}\partial_{r}^{2}\bar{g}_{ab}(x,r_{B})\,,\label{eq:Einsteineq}
    \end{split}
\end{equation}
where we have used Equ.~(\ref{eq:bcbrane}). We can simplify the first equation using Equ.~(\ref{eq:bcbrane}) by noticing that
\begin{equation}
\begin{split}
    \partial_{r}\Big(\text{Tr}(\bar{g}^{-1}\partial_{r}\bar{g})\Big)(x,r_{B})&=-\text{Tr}(\bar{g}^{-1}\partial_{r}\bar{g}\bar{g}^{-1}\partial_{r}\bar{g})+\text{Tr}(\bar{g}^{-1}\partial_{r}^{2}\bar{g})\,,\\&=-8T^{2}+\text{Tr}(\bar{g}^{-1}\partial_{r}^{2}\bar{g})\,,
    \end{split}
\end{equation}
which implies that
\begin{equation}
    \text{Tr}(\bar{g}^{-1}\partial_{r}^{2}\bar{g})(x,r_{B})=4(1+T^2)\,.\label{eq:gprimeprime}
\end{equation}
Taking the trace of the second equation among Equ.~(\ref{eq:Einsteineq}) we have
\begin{equation}
    \bar{R}(x,r_{B})=-2(1-T^{2})\,.\label{eq:Rscalar}
\end{equation}
Moreover, in (2+1)-dimensional situations Equ.~(\ref{eq:gprimeprime}) in fact implies that
\begin{equation}
    \partial_{r}^{2} \bar{g}_{ab}(x,r_{B})\propto \bar{g}_{ab}(x,r_{B})\,,\label{eq:lucky}
\end{equation}
with an $x$-independent proportionality coefficient $2(1+T^2)$.\footnote{This is because of the fact that we are considering pure Einstein gravity in the bulk. This implies that the bulk geometry is always maximally symmetric. Hence the tensor indices of all curvature tensors are carried by the metric.} As a result, we have
\begin{equation}
    \bar{R}_{ab}(x,r_{B})=-(1-T^2)\bar{g}_{ab}\,,\label{eq:Ricci}
\end{equation}
which is the same as Equ.~(\ref{eq:Rscalar}) because in $(1+1)$-dimensions the Einstein tensor is constantly zero. Since the Karch-Randall brane is subcritical, i.e. $\abs{T}<(d-1)$, which for $d=2$ reads $\abs{T}<1$, we have from Equ.~(\ref{eq:Ricci}) that the brane geometry is a hyperbolic Einstein manifold. In our current case the brane is two-dimensional and such a geometry can be constructed as quotient of the upper-half-plane by a Fuchsian group. This proves the existence of the connected phase of the Karch-Randall brane under the constraint of the bulk Einstein's equation.

\subsection{Generalization to BCFT$_{2}$ with Two Boundaries}
The above consideration in this section can be easily generalized to the case of the $\ket{\text{TFD}}$ state of two BCFT$_{2}$'s each with two Cardy boundaries $a$ and $b$ i.e. the case in Fig.\ref{pic:TFD2insertion}. In this case, there are three possible phases of the Karch-Randall branes for the saddle point of the replica gravitational path integral: 1) Both the branes corresponding to the Cardy boundary $a$ and $b$ are disconnected; 2) The brane corresponding to Cardy boundary $a$ is connected and the brane corresponding to Cardy boundary $b$ is disconnected; 3) The brane corresponding to Cardy boundary $a$ is disconnected and the brane corresponding to Cardy boundary $b$ is connected. We note that there is no phase in which both the branes corresponding to the Cardy boundaries $a$ and $b$ are connected as then there is no enough boundary place i.e. $\partial R$ for the bulk $\mathbb{Z}_{n}$ fixed point locus to terminate.\footnote{This is nevertheless possible if the the subregions $R_{I}$ and $R_{II}$ in Fig.\ref{pic:TFD2insertion} are intervals each with two boundaries away from the Cardy boundaries $a$ and $b$.} Hence in the case in Fig.\ref{pic:TFD2insertion} the bulk calculation of $S_{R}$ involves three possible phases of the Ryu-Takayanagi surface: 1) $\gamma_{c}$ connecting the two components of $\partial R$ through the bulk not touching the branes; 2) $\gamma_{dc}^{a}$ connecting $\partial R$ to the brane corresponding the the Cardy boundary $a$; 3) $\gamma_{dc}^{b}$ connecting $\partial R$ to the brane corresponding the the Cardy boundary $b$. As it was shown in detail in \cite{Geng:2021iyq}, this reproduces the answer Equ.~(\ref{eq:CFTEE2}).

\section{Comments on the Higher Dimensional Cases}\label{sec:higherd}
In higher dimensions the replica calculation of the entanglement entropy still follows from Equ.~(\ref{eq:replicatrick}). Nevertheless, due to the lack of local conformal symmetries for higher dimensional CFT's the $\ket{\text{TFD}}$ state cannot be prepared by a path integral on a plane. So the situation will be more complicated.\footnote{Though, the TFD state for which the modular Hamiltonian is a boost operator can be prepared by a path integral on the plane.} However, the essential idea in the above section wouldn't change i.e. the Ryu-Takayanagi surface ending on the brane corresponds to the dominance of the connected phase of the Karch-Randall brane in the replicated gravitational path integral.

Let's take the bulk to be (d+1)-dimensional. To understand the constraint from the bulk Einstein's equation to the brane geometry, let's evaluate the bulk Einstein's equation
\begin{equation}
    R_{\mu\nu}-\frac{1}{2}g_{\mu\nu} R=\frac{d(d-1)}{2}g_{\mu\nu}\,,
\end{equation}
on the brane using the Gauss-Codazzi relations
\begin{equation}
    \bar{R}_{\mu\nu}=R_{\mu\nu}+R_{\rho\sigma}n^{\rho}n^{\sigma}n_{\mu}n_{\nu}-R_{\rho\mu}n^{\rho}n_{\nu}-R_{\rho\nu}n^{\rho}n_{\mu}-R_{\rho\mu\sigma\nu}n^{\rho}n^{\sigma}+K_{\mu\nu}K-K^{\rho}_{\mu}K_{\rho\nu}\,,\label{eq:GC}
\end{equation}
where $\bar{R}_{\mu\nu}$ is the Ricci curvature of the induce metric on the brane and $n^{\rho}$ is the unit normal vector of the brane. As a result, we have
\begin{equation}
\begin{split}
    \bar{R}_{\mu\nu}&=-dg_{\mu\nu}+dn_{\mu}n_{\nu}-R_{\rho\mu\sigma\nu}n^{\rho}n^{\sigma}+K_{\mu\nu}K-K^{\rho}_{\mu}K_{\rho\nu}\,,\\&=-d\bar{g}_{\mu\nu}-R_{\rho\mu\sigma\nu}n^{\rho}n^{\sigma}+K_{\mu\nu}K-K^{\rho}_{\mu}K_{\rho\nu}\,.\label{eq:higherdeom1}
    \end{split}
\end{equation}
Then we can use the brane embedding Equ.~(\ref{eq:braneemb}) to simplify Equ.~(\ref{eq:higherdeom1})\footnote{We should note that we used two different notations for the brane induced metric $\bar{g}_{\mu\nu}$ and $h_{\mu\nu}$.} as
\begin{equation}
\begin{split}
    \bar{R}_{\mu\nu}&=-d\bar{g}_{\mu\nu}-R_{\rho\mu\sigma\nu}n^{\rho}n^{\sigma}+T^{2}\frac{1}{d-1}\bar{g}_{\mu\nu}\,,\\&=-\frac{(d-1)^2-T^2}{d-1}\bar{g}_{\mu\nu}-\bar{g}_{\mu\nu}-R_{\rho\mu\sigma\nu}n^{\rho}n^{\sigma}\,.\label{eq:higherdeom2}
    \end{split}
\end{equation}
where the brane tension is subcritical $\abs{T}<(d-1)$. Let's define the traceless tensor $\bar{g}_{\mu\nu}+R_{\rho\mu\sigma\nu}n^{\rho}n^{\sigma}$ as $T^{m}_{\mu\nu}$, then we see that Equ.~(\ref{eq:higherdeom2}) is in fact the same as Einstein's equation with negative cosmological constant coupled with conformal matter. Moreover, if the bulk geometry is maximally symmetric then $T^{m}_{\mu\nu}=0$ and the brane will be an Einstein manifold. This is consistent with our result in Equ.~(\ref{eq:Ricci}) as when $d=2$ the bulk geometry is always maximally symmetric.

In higher dimensions ($d\geq3$), the Witten-Yau theorem \cite{Witten:1999xp,Cai:1999dqz} says that when $T^{m}_{\mu\nu}=0$ there is no multi-boundary solution of Equ.~(\ref{eq:higherdeom2}) if the asymptotic boundary of the Karch-Randall brane has nonnegative scalar curvature. Hence we see that for a maximally symmetric bulk spacetime the connected phase of the Karch-Randall brane contradicts the bulk Einstein's equation. As a result, in empty AdS$_{d+1}$ bulk one cannot have a brane connecting multiple disconnected boundary submanifolds. This is a no-go theorem to the proposal in \cite{VanRaamsdonk:2020tlr}. 

Nevertheless, in our case we are performing the replica gravitational path integral so the smooth ambient bulk spacetime cannot be maximally symmetric as a smooth maximally symmetric ambient bulk spacetime is necessarily empty AdS$_{d+1}$ which doesn't satisfy the replica boundary condition. Hence we can see that the Witten-Yau theorem doesn't forbid the existence of the connected Karch-Randall brane as a saddle of the replica gravitational path integral.

\section{Discussions and Conclusions}\label{sec:conclusion}
In this paper, we provide a derivation of the Ryu-Takayanagi conjecture in the AdS/BCFT correspondence which was intensively used in the computation of the Page curve and construction of entanglement island in the Karch-Randall braneworld. The derivation translates the replica path integral computation of the entanglement entropy into a bulk gravitational path integral. The bulk contains a Karch-Randall brane. Interestingly, the replica gravitational path integral can be dominated by different phases of the Karch-Randall brane. In general, there are two phases-- the disconnected phase and the connected phase. In the disconnected phase the Karch-Randall brane contains $n$-disconnected components (n is the replica number) and in the connected phase the Karch-Randall brane only has one connected component and it has $n$ disconnected asymptotic boundaries. In the intermediate description of the Karch-Randall braneworld, the disconnected phase is in fact a replica wormhole and it corresponds to the formation of entanglement island on the brane. Moreover, we notice that the connected phase of the Karch-Randall brane as a saddle of the replica gravitational path integral doesn't contradict the Witten-Yau theorem but Witten-Yau theorem does forbid the existence of such connected saddle for the non-replicated gravitational path integral. The later observation provides a no-go theorem for the proposal in \cite{VanRaamsdonk:2020tlr} for a large class of situations. Moreover, we notice that in our case the $n$ boundary replica are in fact coupled to each other as they are glued into a single manifold $\mathcal{M}^{(n)}$ so one does expect the gravitational dual of their boundaries can be a connected geometry i.e. a connected Karch-Randall brane. This is consistent with lessons from string theory \cite{McNamara:2020uza,Heckman:2021vzx} and is opposed to the calculations in \cite{Saad:2019lba,Cotler:2020hgz,Cotler:2020ugk,Cotler:2020lxj,Cotler:2021cqa,Cotler:2022rud} in which bulk geometries connecting disconnected and decoupled asymptotic boundaries are included into the gravitational path integral. We also notice that it is straightforward to extend our work to the cases where the bulk contains higher curvature terms \cite{Dong:2013qoa,Camps:2013zua,Bueno:2020uxs} and the brane contains DGP \cite{Dvali:2000hr,Geng:2023qwm,Geng:2023iqd} and other higher derivative terms \cite{Chen:2020uac}. 

\section*{Acknowledgements}
We are grateful to Amr Ahmadain, Luis Apolo, Andreas Karch, Yikun Jiang and Lisa Randall for relevant discussions. We thank Andreas Karch and Lisa Randall for comments on the draft. HG would like to thank Biwen Tang for the help with pictures in this paper. The work of HG is supported by a grant from Physics Department at Harvard University. 

\appendix
\section{Some Geometric Results} \label{sec:app}
In this appendix, we collect some useful geometric results for the studies in the main text. We are concerned with the geometric properties of following metric
\begin{equation}
ds^2=dr^2+\bar{g}_{ab}(x,r)dx^{a}dx^{b}\,,\quad\text{where $a,b=1,\cdots,d$}\,.
\end{equation}
The nonvanishing components of the Christoffel symbol $\Gamma^{\mu}_{\nu\rho}$ are
\begin{equation}
    \Gamma^{r}_{ab}=-\frac{1}{2}\partial_{r}\bar{g}_{ab}\,,\quad \Gamma^{a}_{rb}=\Gamma^{a}_{br}=\frac{1}{2}\bar{g}^{ac}\partial_{r}\bar{g}_{bc}\,,\quad \Gamma^{a}_{bc}=\bar{\Gamma}^{a}_{bc}\,,
\end{equation}
where $\bar{\Gamma}^{a}_{bc}$ is the Christoffel symbol of $\bar{g}_{ab}$. The nonvanishing components of the extrinsic curvature of the constant-$r$ slices are
\begin{equation}
    K_{ab}=\frac{1}{2}\partial_{r} \bar{g}_{ab}\,.
\end{equation}
Hence we have
\begin{equation}
    \Gamma^{r}_{ab}=- K_{ab}\,,\quad \Gamma^{a}_{rb}=\Gamma^{a}_{br}= K^{a}_{b}\,.
\end{equation}
The components of the Ricci tensor are
\begin{equation}
    \begin{split}
        R_{rr}&=-\partial_{r} K- K^{ab}K_{ab}\,,\\ R_{ra}&=R_{ar}=\bar{\nabla}_{b} K^{b}_{a}-\partial_{a} K\,,\\ R_{ab}&=\bar{R}_{ab}-\partial_{r}K_{ab}-KK_{ab}+2K_{ac}K^{c}_{b}\,.
    \end{split}
\end{equation}
\pagebreak
\bibliographystyle{JHEP}

\bibliography{main}

\end{document}